# Analytical and empirical measurement of fiber photometry signal volume in brain tissue

Marco Pisanello[1, +], Filippo Pisano[1, +], Minsuk Hyun[2, +], Emanuela Maglie[1, 3], Antonio Balena[1, 3], Massimo De Vittorio[1, 3], Bernardo L. Sabatini[2, †, *], Ferruccio Pisanello[1, †, *]

1. Istituto Italiano di Tecnologia, Center for Biomolecular Nanotechnologies, 73010 Arnesano (LE), Italy.
2. Department of Neurobiology, Howard Hughes Medical Institute, Harvard Medical School, 02115 Boston (MA), U.S.A.
3. Dipartimento di Ingegneria dell'Innovazione, Università del Salento, 73100 Lecce (LE), Italy.

+ These authors equally contributed to this work.
† These authors equally contributed to this work.
* Corresponding authors: bernardo_sabatini@hms.harvard.edu - ferruccio.pisanello@iit.it

**Abstract**

Fiber photometry permits monitoring fluorescent indicators of neural activity in behaving animals. Optical fibers are typically used to excite and collect fluorescence from genetically-encoded calcium indicators expressed by a subset of neurons in a circuit of interest. However, a quantitative understanding of the brain volumes from which signal is collected and how this depends on the properties of the optical fibers are lacking. Here we analytically model and experimentally measure the light emission and collection fields for optical fibers in solution and scattering tissue, providing a comprehensive characterization of fibers commonly employed for fiber photometry. Since photometry signals depend on both excitation and collection efficiency, a combined confocal/2-photon microscope was developed to evaluate these parameters independently. We find that the 80% of the effective signal arises from a $10^5$-$10^6$ $\mu m^3$ volume extending ~200 μm from the fiber face, and thus permitting a spatial interpretation of measurements made with fiber photometry.

## 1. Introduction

In the last decade, optogenetics has become widely used for optical control of neural activity [1–3]. Simultaneously, new implantable devices, mainly based on waveguides [4–11] or micro light emitting diodes (μLEDs) [12–17], have been developed for light delivery in the living brain. Recently these optical approaches have been extended to also monitor neural activity, detecting fluorescence variations over time of genetically-encoded fluorescent indicators of intracellular calcium (GCaMP) or transmembrane voltage [18–23]. Although new devices utilizing integrated photodetectors and μLEDs have been described [24], traditional flat-cleaved optical fibers are broadly used for both triggering and collecting fluorescence *in vivo* in freely behaving animals. This approach is typically referred to as fiber photometry [25–44].

While emission properties and light delivery geometries of optical fibers in the brain are well-known [45–47], the study of their light collection properties is lagging. Even though an analytical model in quasi-transparent medium can be derived [48], only Monte Carlo simulations with spatial resolution of tens of micrometers have been used to estimate light collection properties from volume containing turbid medium [49–53]. Experimental measurements have been done mainly to evaluate on-axis performance (i.e. how light collection efficiency depends on the distance from the fiber face along the main axis of the fiber) [54]. However, *in vivo* much of the light collected likely both arises from off-axis flourophores and depends on the numerical aperture and size of the fiber in a manner that is fundamentally different. Thus, an overall understanding of volumetric light collection is lacking, and is needed to help neuroscientists in designing experiments and interpreting their data.

In this study, we characterize the light collection properties of optical fibers typically used to monitor fluorescence in the brain. A combined confocal/2-photon laser-scanning microscope is used to measure the light collection and emission diagrams in scattering tissue. The two-photon path was used to raster scan a point-like fluorescence source around the fiber facet. Measurement of the fluorescence collected by the fiber as the point source moves provides a measurement of the 3D collection maps. In addition, descanned pinhole detection, implemented on the same scanhead, was used to measure emission diagrams within the field of view, thus providing an understanding of the distribution in the tissue of excitation photons emitted by the fiber. The two information are combined to determine the "*photometry efficiency – $\rho(x,y,z)$*", taking into account both excitation and collection efficiency in the tissue volume of interest. This method is exploited to compare flat-cleaved optical fibers with different numerical aperture (NA) and core diameter (*a*) in semi-transparent solutions and in brain slices. We found that fibers with different NA and the same core diameter behave similarly in terms of spatial decay along the fiber axis, with collection volumes larger for higher NA mostly due to out-of-axis signal. These findings are supported by analytical and ray tracing collection maps. Our results highlight previously unknown aspects of light collection from brain tissue with optical fibers that will guide the design of future fiber photometry experiments.

## 2. Results

*2.1 Numerical estimation of optical fibers collection field*

As schematically represented in Fig. 1A, the light generated from an isotropic fluorescent source is collected by an optical fiber with a certain efficiency that depends on optical fiber properties (numerical aperture and diameter) and on the properties of the medium between the source and the fiber (refractive index, absorption and scattering). For a given fiber with core diameter *a* and numerical aperture NA, immersed in a homogeneous medium with refractive index *n*, the analytical approach provided by Engelbrecht *et al.* [48] estimates the 2D map of collection efficiency $\psi$ (NA, *n*, *a*, *x*, *z*) as the fraction of the power collected by the fiber core from an isotropic point source located in the (*x*, *z*) place (see Fig. 1A for axis definition). To numerically estimate the collection field of optical fibers typically employed for *in vivo* fiber photometry, while considering tissue absorption and scattering, we combined the approach in Ref. [48] with a ray tracing model. In the following, we first extend the method proposed by Engelbrecth *et al.* [48] to take into account the light entering the waveguide from the cladding front face; we then use the results to validate a ray tracing model that numerically estimates the fiber collection field in scattering brain tissue (see Materials and Methods).

The collection efficiency $\eta$ for a fiber with core diameter *a*, cladding diameter *b*, core refractive index $n_{core}$, and cladding refractive index $n_{clad}$ can be written as

$$\eta(\text{NA}, n_{\text{core}}, n, a, b, x, z) = \psi(\text{NA}, n, a, x, z) + \psi(\text{NA}_{\text{eq}}, n, b, x, z) - \psi(\text{NA}_{\text{eq}}, n, a, x, z), \quad (1)$$

where $\text{NA}_{\text{eq}} = \sqrt{n_{\text{clad}}^2 - n^2} = \sqrt{n_{\text{core}}^2 - \text{NA}^2 - n^2}$ is the equivalent numerical aperture of the cladding/external medium interface [55], with the term $\psi$ (NA$_{eq}$, *n*, *b*, *x*, *z*) - $\psi$ (NA$_{eq}$, *n*, *a*, *x*, *z*) accounting for light collected by the cladding. Taking advantage of the cylindrical symmetry of the system, values of $\eta$ throughout the whole space filled by the external medium can be obtained. Meridional slices ($y = 0$) of such volumes are shown in Fig. 1B for optical fibers with three different configurations of NA/core diameter (0.22/50μm, 0.39/200μm and 0.50/200μm, respectively). These maps show the presence of a region with constant collection efficiency next to the fiber core, roughly described by a cone with base coincident with the fiber facet surface and vertex lying on the waveguide axis at [48]

$$z_0 = \frac{a}{2 \cdot \tan(\text{NA}/n)}. \quad (2)$$

Interestingly, for 0.39/200μm fiber the maximum collection efficiency region lies along the lateral surface of the cone, due to the fact that $\text{NA}_{eq} > \text{NA}$, whereas this does not happen in the case of the 0.50/200μm fiber. This difference between 0.39/200μm and 0.50/200μm fibers is clearly visible also in the axial collection profiles (blue lines) (Fig. 1E). In the case of the 0.39/200μm fiber, as a result of cladding collection a peak in the collection efficiency is observed at the boundary of the constant region close to the fiber (red arrow in Fig. 1E, middle panel) on the other hand, for the 0.50/200μm cladding collection is reflected only in a local change of slope (red arrow in Fig. 1E. bottom panel).

Ray-tracing simulations were performed by scanning an isotropic point source at $\lambda = 520$nm across the $xz$ plane (the ray tracing setup is shown in Supp. Fig. S1). Modeled light rays entering the fiber within $\text{NA}_{eq}$ were collected through a short length of patch fiber (200mm) and measured if they reached a hypothetical detector at the distal end of the fiber (Fig. 1C). This configuration simulated the potential leakage of light rays outside $\text{NA}_{eq}$ that propagate in the cladding. A comparison in terms of axial collection profiles in (Fig. 1E) shows a very good agreement with the analytical model for both the geometrical behavior and the absolute collection efficiency values. In particular, the maximum collection efficiency $\eta$ for the 0.39/200μmNA fiber was found to be ~ 0.03 whereas in the case 0.50/200μm it was estimated to be ~ 0.04.

Since the ray-tracing approach gave results consistent with those derived with the analytical method, we extended the numerical simulations to model turbid media, such as scattering brain tissue. We modeled the medium around the fiber with Henyey-Greenstein scattering, to simulate absorption and scattering properties of brain tissue [56, 57] (refractive index $n = 1.360$, mean free path $l = 48.95$μm, anisotropy parameter $g = 0.9254$, transmission coefficient $T = 0.9989$). The resulting maps of collection efficiency for 0.22/50μm, 0.39/200μm, and 0.50/200μm fibers are shown in Fig. 1D. As a comparison, the axial profiles of analytical and numerical estimation of $\eta$ for both transparent and turbid media are reported in Fig. 1E for all the investigated fibers. When absorption and scattering of the medium are taken into account, the constant region in collection efficiency almost disappears, and $\eta$ starts decreasing immediately after the fiber facet. The maximum $\eta$ value remains ~ 65% higher for the 0.50/200μm fiber with respect to 0.39/200μm. However, the collection efficiency decrease is slightly steeper for the 0.50/200μm fiber, reaching 50% of the maximum at 250μm from the fiber facet, compared to 300μm observed for the 0.39/200μm fiber.

*2.2 Direct measurement of collection field in quasi-transparent fluorescent media*

A two-photon (2P) laser scanning system has been designed and built to directly measure the light collection field of optical fibers. A block diagram of the optical path is illustrated in Fig. 2A: the optical fiber was submerged in a fluorescent PBS:fluorescein solution (30μM) and a fs-pulsed near-infrared laser ($\lambda_{ex}$ = 920nm) is used to generate a fluorescent voxel that is scanned in three dimensions close to the fiber facet. Scan in the *xz* plane was obtained by a two-axis galvanometric scanhead on a ~1.4×1.4mm² field of view (FOV), with the microscope objective (Olympus XLFluor 4x/340) mounted on a *y*-axis piezo focuser to obtain volumetric scan. The voxel emission is collected by the same objective and detected by a non-descanned photomultiplier tube (*"μscope PMT"*). This gives a measurement of the total fluorescence generated and, if needed, can be used to compensate for changes in excitation efficient by the scanning point source. Simultaneously, the fraction of the voxel's fluorescence that is collected by the optical fiber and guided to a second PMT (*"fiber PMT"*) was measured. The point spread function (PSF) of the two-photon epifluorescence system has been measured to be 3μm laterally and 32μm axially (see Supp. Fig. S2 and Materials and Methods for details), with the large axial extent resulting for the low NA objective necessary to scan over a large field of view.

During volumetric raster scanning of the 2P spot, Scanimage software (Vidrio Technologies) was used to reconstruct images from both the *μscope PMT* and the *fiber PMT* signals, allowing point-by-point mapping of the light intensity collected from the optical fiber within the scanned volume. Fig. 2B shows the signal collected by the *fiber PMT* when the excitation is scanned across the *y* = 0 plane for three different type of optical fiber: 0.22/50μm (Thorlabs FG050UGA, top panel), 0.39/200μm (Thorlabs FT200UMT, middle panel), and 0.50/200μm (Thorlabs FP200URT, bottom panel), with overlay of the isolines at 10%, 20%, 40%, 60%, and 80% of the maximum number of photon collected. Supp. Fig. 3 shows the signal collected along the *x* = 0 plane. These images were corrected for unevenness of illumination within the FOV by using the related signal on the *μscope PMT*. In addition, the gain *G* of the system was estimated by noise analysis at each measurement session and used to convert the PMT signals into numbers of photons (see Materials and Methods for details). For a direct comparison, analytical collection maps were also computed convolving $\eta$ and a three-dimensional function modeling the experimental PSF (Fig. 2C) (details on this calculation are reported in Materials and Methods). The related axial profiles, shown in Fig. 2D, indicate good agreement between numerical predictions and the experimental data. Both analytical and experimental results show a difference between 0.39/200μm and 0.50/200μm fibers. For 0.39/200μm fibers (Fig. 2B), the region of maximum collection does not lie on the optical axis, but in two lobes near the boundary of the core: this can be ascribed to the fact that light is efficiently guided not only by the core-cladding interface, but also by the waveguide formed by the cladding and the external medium (*i.e.* the PBS:fluorescein solution).

By assembling the data from the measurement planes collected into the volume, the full three-dimensional collection fields can be reconstructed and used to illustrate the iso-intensity surface at 10%, 20%, 40%, 60%, and 80% of the maximum number of collected photons (Fig. 3A). The volumes enclosed by these surfaces reflect those from which a given fraction of the collected photons arise and hence determine the effective volume from which functional signals can be detected (Fig. 3B). Collection volumes of 0.39NA/200μm and 0.50NA/200μm fibers behave very similar for relative intensities <=60%, with the 0.39NA/200μm collecting with the highest intensity from a smaller volume with respect to the 0.50/200μm fiber. The 0.22NA/50μm (which has a 16 times smaller core surface) shows consistently lower collection volumes. When the volumetric data for the 0.22NA/50μm fiber are multiplied by a factor 16 (dotted yellow line in Fig. 3(b)), collection volumes are very close to those of 0.39/200μm and 0.50/200μm fibers. In addition, the absolute value of collected photons, shown in Fig. 3C, increases as a function of NA, allowing for measurements with a higher signal-to-noise ratio.

*2.3 Direct measurement of collection field in brain slices*

The system depicted in Fig. 2A was also used to measure the light collection field of 0.39/200µm and 0.50/200µm optical fibers in 300µm thick brain slices stained with fluorescein. This was done to estimate the influence of tissue absorption and scattering on the geometrical features of light collection. Fig. 4A shows the results of these measurements for the two investigated fibers on the plane $y = 0$, with overlay of the isolines at 10%, 20%, 40%, 60%, and 80% of the maximum signal detected from the *fiber PMT*. A clear difference compared to the measurement in PBS:fluorescein solutions is that the flat collection efficiency region was disrupted by tissue scattering such that it was not possible to define the characteristic point at $z_0$ for either fiber. This is also seen in the axial collection profiles reported in Fig. 4B, which show a very steep decrease of the collection curve starting at the fiber face. These findings are confirmed by comparing the results to those obtained using the ray-tracing model discussed above for both axial collection profiles (red lines in Fig. 4B) and their derivatives (Supp. Fig. S4A).

Taking advantage of the cylindrical symmetry of the fiber properties, the $y = 0$, $x < 0$ half-planes of the measurements in Fig 4A were used to reconstruct the collection volume. The collection volume was reconstructed by 360° rotation after applying a 11x11 pixel moving average filter (details on the procedure are given in Materials and Methods). Iso-intensity surfaces of the reconstructed 3D collection field at 10%, 20%, 40%, 60%, and 80% of the maximum PMT counts value were calculated (Fig. 4C). From these, their enclosed volume was measured and compared to those obtained in solution (Fig. 4D). The collection volumes in tissue are clearly smaller with respect to the fluorescent solution driven as expected from light absorption and scattering. In addition, the higher numerical aperture of the 0.50/200µm fiber increased the average collection volumes by a factor ~1.8 compared to the 0.39/200µm volumes (See Supp. Fig. S4B for detailed plot of volumes ratio between 0.50 and 0.39 fibers for each iso-intensity curve).

*2.4 Photometry efficiency measurement*

Above we describe an approach to estimate the collection efficiency of an optical fiber by scanning a point like source in the proximity of the fiber facet. However, in fiber photometry experiments fluorescence is generated by delivering excitation light (typically at 473nm or 488nm) and collecting the generated fluorescence (usually in the range 500nm-550nm) through the same fiber. Therefore, light intensity obtained from a specific position depends not only on how photons are collected from that point, but also on the efficiency at which fluorescence is excited at that point. A *photometry efficiency* parameter $\rho$ can therefore be defined as:

$$\rho(x,y,z) = \eta(x,y,z) \cdot \beta(x,y,z), \quad (3)$$

where $\eta$ is the collection efficiency and $\beta$ is the normalized light emission diagram of the same optical fiber used to collect light. In brain tissue, $\eta$ can be estimated with the 2P scanning method detailed in paragraph 2.3. To estimate $\beta$ in the same location where $\eta$ is measured, pinhole detection was implemented in a de-scanned collection path. The scanning pinhole allows light to reach the detector only if it arises from a conjugate location in the tissue; thus, its intensity is determined by the efficiency with which light reaches the point from the fiber. In this way fiber emission diagram is measured in brain slices uniformly stained with fluorescein. A block diagram of the experimental setup is shown in Fig. 5A. A 473nm CW laser source is coupled to the fiber distal end and provides the excitation light. The fluorescence light excited in the brain slice by the optical fiber is imaged on the galvanometric mirrors and scanned through a pinhole aperture that conveys it on a *pinhole PMT*. This detector is used to reconstruct the fluorescence intensity map within the same FOV (same magnification and position) of the 2P scan used to estimate $\eta$ (Fig. 5B-C, respectively.) These maps, upon normalization, give an estimation of the light emission diagram $\beta$ and can be used in conjunction with the collection fields to estimate $\rho$ (pixel by pixel product of $\eta$ and $\beta$).

The resulting maps (Fig. 5D, with overlay of the isolines at 10%, 20%, 40%, 60%, and 80% of the maximum photometry efficiency) contain in each pixel a value proportional to *(i)* the amount of fluorescence light excited by the fiber in that pixel and *(ii)* to the amount of light collected from that pixel. These values describe the relative contribution of signal arising from each voxel and thus determine the spatial distribution of the sources of signal collected during a fiber photometry recording. This analysis reveals that, as expected from the effect of tissue scattering, the axial profile of $\rho$ falls of faster with distance from the fiber face than the collection only diagram (Fig. 5E) for both the 0.39/200μm and 0.50/200μm fibers.

Similarly, volumetric analysis analogous to the ones described in Section 2.3 can be extended to photometry efficiency (Fig. 6) to determine the volumes enclosed by the iso-intensity surfaces at 10%, 20%, 40%, 60%, and 80% total photometry efficiency. The higher numerical aperture 0.50/200μm fiber results in a normalized collection volume ~2.0 times higher than the 0.39/200μm fiber (on average: the plot of the ratio between the two datasets for the different iso-intensity surfaces is reported in Supp. Fig. S5).

## 3. Discussion

Although fiber photometry is regularly employed to investigate the relationship between neural activity and behavior as well as connectivity in neural circuits [25, 26, 37–44], the collection properties of optical fibers inserted into the brain are not yet well characterized. Neither analytical methods nor experimental data exist describing the spatial domain over which signals are collected and how it depends on the properties of the optical fibers used. In this work we introduce a series of approaches that define a photometry efficiency parameter. This is achieved by implementing a combined confocal/two-photon laser-scanning microscope to measure both the collection and emission diagrams from the same fiber in the same region (Fig. 2 and 4). The 2P path was used to estimate light detection diagrams in both quasi-transparent media and brain slices (Fig. 2 and 3, respectively), by generating a point-like source raster scanned close to the fiber facet. Fluorescence light is synchronously collected by the fiber, giving access to the three-dimensional collection map. The resulting data consistently agree with analytical and numerical estimations based on theory and ray tracing models. Even though the results shown in this work are referred to cortical area with fluorescence detection at 500-550nm, the method is readily adaptable to any wavelength and applicable to any brain region, potentially allowing to map brain regions that are better suited for collecting functional fluorescence. The method can also be applied to evaluate fiber photometry performances in any fluorescent turbid media.

According to our data, when scattering effects are neglected (*i.e.* the fiber is immersed in a homogeneous, non-turbid medium), the dominant effects of fiber size and numerical aperture can be evaluated. We find that the collection volumes mostly depend on the fiber core size, $a$, whereas NA affects the total number of photons collected. From a practical point of view, we formulate a photometry efficiency parameter, $\rho$, taking into account the overlap between the light collection fields and the light delivery diagrams. The latter was evaluated in stained brain slices (Fig. 5) through the confocal path.

In this respect, 0.39/200μm and 0.50/200μm fibers behave very similarly in terms of on-axis spatial decay of the collected signals (Fig. 5E). The 0.50/200μm fiber, however, collects signal from a volume approximately two times bigger with respect to the 0.39/200μm fiber (Fig. 4D and 6B), mainly due to out-of-axis contributions. Our results suggest that the influence of the numerical aperture in defining the axial (dorsal-ventral with a typical implantation of a fiber laong this axis) extension of the brain volume under investigation is marginal with respect to the effect of the fiber diameter. Optical fibers with smaller core, such as the 0.22/50μm one, can be utilized to collect functional fluorescence signals from a restricted tissue volume, when localized information is needed.

From a practical perspective of fiber photometry experiments, data reported here help to set criteria in choosing the optical fiber to use in a specific experiment: (i) the fiber core size should be chosen according to the volumetric extent of the functional area of interest; (ii) if on-axis contribution is preferred (e.g. cortical columns), a lower NA should be used; (iii) if a low count rate is expected, high NA fibers can help by detecting additional out-of-axis fluorescence.

We propose that the methods used here can become part of photometry measurements pipelines, especially when unconventional devices are used (since it can be simply extended to any fiber-coupled device): scientists could potentially benefit from its application in the selection of the best device to match their experimental needs. Moreover, the visualization of the effective area involved in the measurement allows for the individuation of sub-regions of interest, refining both experimental design and data analysis.

**Materials and Methods**

*Fiber stubs fabrication*

Fibers stubs were realized from 0.22/50μm (Thorlabs FG050UGA), 0.39/200μm (Thorlabs FT200UMT), and 0.50/200μm (Thorlabs FP200URT) multimode optical with cladding diameters of 125μm and 225μm, for 0.22/50μm and 0.39/200μm-0.50/200μm fibers respectively. After peeling off the buffer, stubs were trimmed to size using a fiber cleaver (Fujikura CT-101) and connectorized to a stainless-steel 1.25mm ferrule. Metallic ferrules were employed to minimize auto-fluorescence artifacts. The connectorized ends of the stubs were then manually polished. Fiber patches were realized from the same fiber types and connectorized using a stainless-steel ferrule on the proximal end and a SMA connector on the distal end, with respect to the stub. During experiments, the stubs were connected to a patch fiber of matching NA and core/cladding sizes via ferrule to ferrule butt-coupling. The patch fiber has a fully preserved core/cladding/buffer structure, that influence propagation of light collected and guided by the cladding of the fiber stub. For the 0.39/200μm and 0.50/200μm fibers, light entering the cladding can propagate in the patch fiber since the buffer has a refractive index $n_{buf} < n_{clad}$. For 0.22/50μm fiber, instead, the acrylate buffer refractive index is $n_{buf}$ = 1.4950 at 520nm [58], higher than $n_{clad}$ = 1.4440, thus preventing light to be guided into the cladding of the patch fiber.

*Analytical calculation of fiber collection efficiency*

Analytical 3D maps of fiber collection fields for an ideal point source were calculated for 0.22/50μm, 0.39/200μm, and 0.50/200μm fibers following Engelbrecht *et al.* [48]. We extended their work to include the cladding front face contribution to the collection fields of 0.39/200μm and 0.50/200μm fibers by implementing Eq. (1). The refractive index of the silica core ($n_{core}$ = 1.4650 at 520nm) has been retrieved from the online database [59], while the refractive index of the cladding ($n_{clad}$) has been calculated to provide the nominal numerical aperture. In Eq. (1), $\psi$(NA, n, a, x, z) is the same as in Ref. [48], while the term $\psi$(NA$_{eq}$, n, b, x, z) - $\psi$(NA$_{eq}$, n, a, x, z) takes into account light collection from the cladding facet (see Supp. Script 1). In particular, $\psi$(NA$_{eq}$, n, b, x, z) is the collection efficiency of a fiber with diameter b that guides light by virtue of the refractive index contrast between the surrounding medium and the cladding with numerical aperture NA$_{eq}$. However, the cladding has a thickness b-a with circular crown shape, and therefore the collection efficiency from the region overlapped to the core, $\psi$(NA$_{eq}$, n, a, x, z), should be subtracted. This modeling was used for obtaining the collection efficiency maps for point-like sources shown in Fig 1(b). To extend these data to the case of an extended source, three dimensional maps for point-like source were obtained by rotating collection efficiency maps in the y = 0 plane around the fiber axis (see Supp. Script 1). The obtained 3D maps were then convolved with a 3D representation of the actual focal spot generated by the microscope (see Supp. Script 1). This latter was modeled as a Gaussian function with lateral FWHM $r_{x,z}$ = 3μm, and axial FWHM $r_y$ = 32μm, modeling the PSF of the used experimental configuration (see below for details in the PSF measurements).

*Ray tracing simulation of fiber collection efficiency*

Ray tracing simulations were performed using an optical model designed with commercial optical ray-tracing software Zemax-OpticStudio to simulate the behavior of light collected by the optical fibers. The implemented layout is shown in Supp. Fig. S1. Flat fibers were represented as two nested cylinders simulating core and cladding of nominal diameters (50μm/125μm for 0.22NA Thorlabs FG050UGA fiber, 200μm/225μm for 0.39NA Thorlabs FT200UMT and 0.50NA Thorlabs FP200URT fibers). Numerical apertures of the fibers were defined by setting the respective core/cladding refractive indexes $n_{core}/n_{clad}$ as 1.4613/1.4440 for 0.22/50μm fiber, 1.4613/1.4079 for 0.39/200μm fiber and 1.4613/1.3730 for 0.50/200μm fiber [59]. One of the two fiber facets was included within an optically homogeneous cylinder volume that simulated the PBS:fluorescein droplet or the stained brain slice. A fluorescence source was modeled as a 520 nm point source emitting 500k rays for a total power of 1W. To reproduce the experimental acquisition, the source was raster scanned across the half-plane y = 0, x > 0 (Fig 1(a)). To optimize simulation time, steps along z and x were set to 25μm; for simulations concerning 0.22/50μm fiber the region in the proximity of the fiber (600μm along z and 500μm along x) were simulated with a grid step of 12.5μm, to better sample the smaller core. For each source position, the rays were collected from both core and cladding surfaces on the facet, propagated into the fiber and they were detected on a single-pixel squared detector placed at the distal end of the fiber. The detector size was matched to the cladding diameter. Refractive indexes were set as n = 1.335 for PBS:fluorescein solution and as 1.360 for brain-like scattering volume [60]. Scattering in the PBS:fluorescein solution was not modeled. Scattering in brain tissue was simulated following a Henyey-Greenstein model with parameters: mean free path m = 0.04895mm, anisotropy value g = 0.9254 and transmission T = 0.9989 [56, 57].

From the computational point of view, the most demanding part of the simulation is rays propagation into the fiber, that experimentally is ~1m long and requires > 24h per simulation. To shorten simulation times, a relatively short length of the fibers was implemented (200 mm). This short length does not allow, however, taking into account losses of light entering the fiber outside the maximum accepted angle. Therefore, only rays describing an angle with the fiber input facet smaller than a threshold $\theta_{\text{th}}$ were considered in the power count, with

$$\theta_{\text{th}} = \sin^{-1}\frac{\max\{\text{NA}, \text{NA}_{\text{eq}}\}}{n}. \quad (4)$$

For the 0.22/50μm fiber the cladding sidewalls were modeled as an absorbing interface to take into account for the leakage of light from the cladding ($n_{\text{clad}}$ = 1.4440 at 520nm [58]) to the buffer ($n_{\text{buf}}$ = 1.4950 at 520nm [58]) into the patch fiber.

*Brain slices treatment*

Brain slices ~300μm thick were cut with a vibratome from wild-type mice brain. Slices were then fixed in PFA and permeabilized for 30min in 0.3% Triton X-100 (Sigma-Aldrich). Slices were then incubated with fluorescein (1mM in PBS) for 30min.

*Acquisition and analysis of fiber collection fields*

A combined confocal/two-photon laser scanning microscope was designed and built to perform the optical characterization proposed in this work. A full block diagram of the path used to measure collection fields is illustrated in Fig. 2(a). The power of a fs-pulsed near-infrared (NIR) laser beam (Coherent Chameleon Discovery, emission tuned at $\lambda_{\text{ex}}$ = 920nm) is modulated by means of a Pockels cell (Conoptics 350-80-02), and a quarter wave plate (Thorlabs AQWP05M-980) has been used to obtain circularly polarized light. The laser beam is expanded by a factor 5 and *xz*-scanned with a galvo/galvo head (Sutter). The microscope objective (Olympus XLFluor 4x/340) is mounted on a *y*-axis piezo focuser (Phisik Instrument P-725.4CD), and fluorescence signal is excited into a quasi-transparent 30μM PBS:Fluorescein solution or into a fluorescently stained brain slice. Fluorescence light is re-collected by the same objective and conveyed without descanning on the entrance window of a photomultiplier tube (PMT, Hamamatsu H10770PA-40, the "μscope PMT") through a dichroic mirror (Semrock FF665-Di02), two spherical lenses (Thorlabs LA1708-A and LA1805-A), and a bandpass filter (BPF, Semrock FF01-520/70-25). During experiments in solution, fiber stubs were submerged in a PBS:Fluorescein droplet held in the sample plane by a hydrophobic layer. After a butt-to-butt coupling with a patch fiber of matched NA and core/cladding diameter, the light back emitted from the fiber was collected through a microscope objective (Olympus Plan N 40x) and sent to the entrance window of a PMT (Hamamatsu H7422P-40, the "fiber PMT"), through two spherical lenses (Thorlabs LA1050-A and LA1805-A) and a BPF (Semrock FF03-525/50-25).

A focal spot was then generated and scanned in the vicinity of the fiber facet covering a field of view of ~1.4×1.4mm² with 512×512 pixels, with the beam resting on each pixel for ~3.2μs. Laser power and PMTs gain were adjusted to optimize signal to noise ratio. For each measurement, a 400μm thick stack was acquired with a 5μm step along *y*, starting slightly below the fiber axis and finishing above the fiber. Each slice in the stack was averaged out of 5 frames.

The number of photons for each frame was calculated as $N_{\text{ph}} = PMT_{\text{counts}}/G$, where $G$ represents the gain of the acquisition system. $G$ was measured as $G = \sigma^2_{\text{counts}}/\langle PMT_{\text{counts}}\rangle$, where the average number of counts $\langle PMT_{\text{counts}}\rangle$ and the variance $\sigma^2_{\text{counts}}$ were acquired by illuminating a confined and homogeneous region in the fluorescein drop. Stacks acquired through the fiber PMT were corrected slice-by-slice for unevenness in excitation, by scaling them against the normalized corresponding image collected by the μscope PMT. The frame acquired for gain measurement of the epi-fluorescence path has been used to correct for slight variability of laser power between measurements, proportionally to the pixel average value. Uncertainty $\sigma_c$ on the cumulative number of photons shown in Fig. 3(c) were evaluated propagating the Poisson noise on the photon count of every pixel and the error on gain determination as

$$\sigma_c(z) = \sqrt{\sum_{x,y,z}\left\{N_{ph}(x,y,z)\sqrt{\left(\frac{1}{\sqrt{PMT^{fiber}_{counts}(x,y,z)}}\right)^2 + \left(\frac{1}{\sqrt{PMT^{\mu scope}_{counts}(x,y,z)}}\right)^2 + \left[\frac{std(G^{fiber})}{\langle G^{fiber}\rangle}\right]^2}\right\}} \quad (5)$$

where the superscript *fiber* and *μscope* identify the PMT, mean and standard deviation of *G* are evaluated over five consecutive frames, and the sum indexes span across the whole *xy* plane and up to *z*. The value of $\sigma_c$ resulted to be less than 1% of $N_c$ at all *z* for all fibers. Data processing was done through Supp. Script 2 and Supp. Script 3 for images collected in quasi-transparent medium and in brain slice, respectively.

*Point spread function measurement*

The PSF of the two-photon microscope was measured by imaging sub-resolution nanoparticles (100 nm) at 920nm with 160nm lateral steps and 2μm axial steps. For the 4X/0.28NA Olympus XLFluor 4x/340 objective, this resulted in a PSF with lateral FWHM $r_{x,z}$ = 3μm ± 1μm, and axial FWHM $r_y$ = 32μm ± 5μm (Supp. Fig. S2). Lateral and axial profiles were fitted with a gaussian function. Two nanoparticles were considered in this measurement, values shown are mean ± standard deviation.

*Acquisition of spatially sampled fiber emission diagrams*

The setup schematically shown in Fig. 5(a) was used to measure the emission diagrams of flat-cleaved optical fibers in tissue. Fibers were inserted into a 300μm thick fluorescently stained brain slice, 473nm light was coupled into the fiber through an objective lens (Olympus Plan N 40x), and the primary dichroic of the 2P microscope was removed from the system. Light emission from the tissue was collected through the microscope objective, descanned by the scan-head, focused into a pinhole (Thorlabs MPH-16), and detected by a PMT (Hamamatsu H7422P-40, the "*pinhole PMT*"). A BPF (Thorlabs MF525/39) isolated the wavelength band of interest. The pinhole size was set to 100μm.

*Photometry efficiency calculation*

Images acquired on the *y* = 0 plane by the *fiber PMT* and the *pinhole PMT* were used to determine the photometry efficiency. The image acquired through the pinhole was registered over the collection field to obtain a pixel-to-pixel spatial correspondence. The photometry efficiency maps were determined as the pixel by pixel product of normalized version of illumination and collection fields (see Supp. Script 4). One half of the photometry efficiency maps was employed to obtain a volumetric representation of this quantity, as reported previously.

*Matlab programming*

Data processing was implemented in Matlab. Scripts are reported in supplementary materials, with Supp. Script 5 containing all the function called in Supp. Script 1-4 without definition.

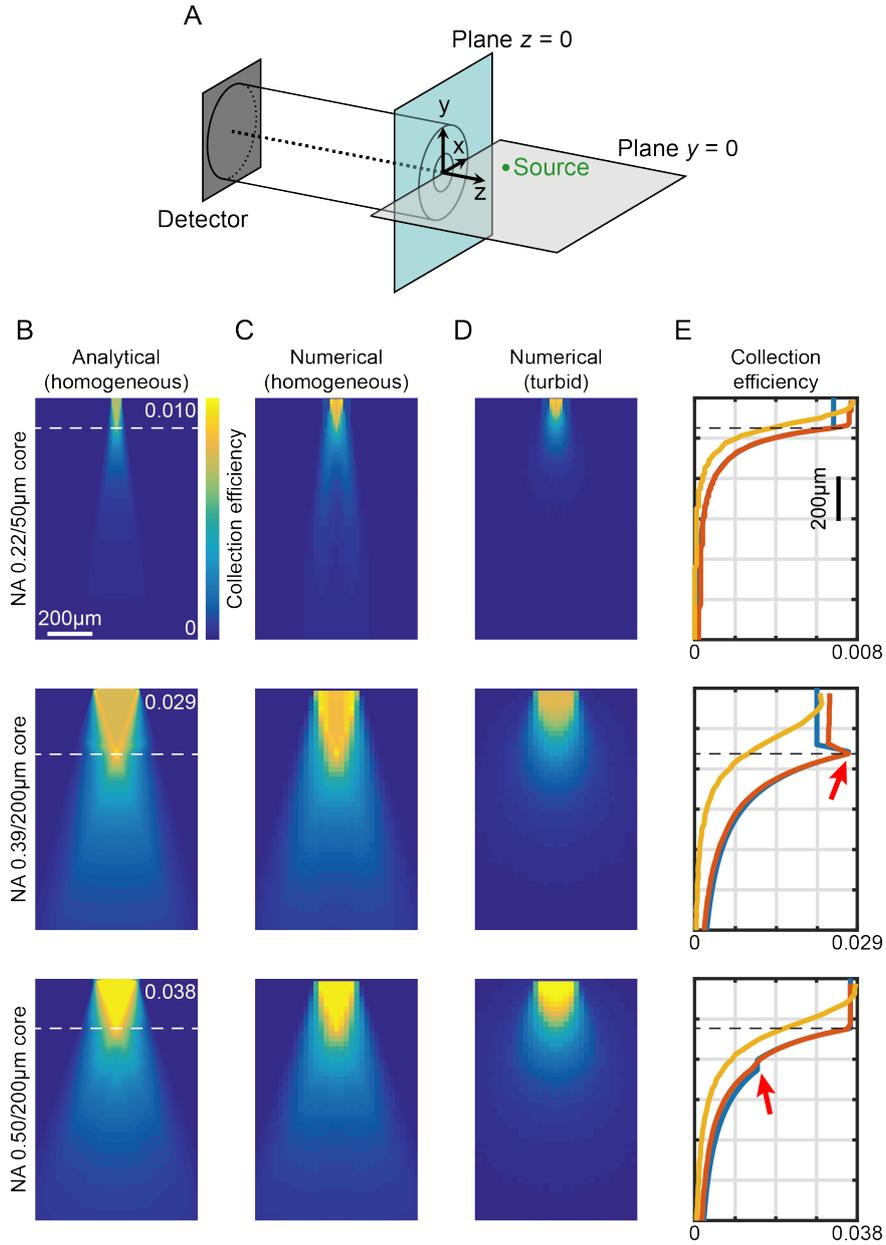

**Figure 1. Computation model of light collection efficiency for optical fibers**

**A,** Reference system used throughout the manuscript. An example point source (green) is shown in the plane $y = 0$.

**B,** Analytical calculations of collection efficiency diagrams for light emitting from point sources locating in the $xz$ ($y = 0$) plane for three different fibers. Data are shown for 0.22NA/50μm, 0.39NA/200μm, and 0.50NA/200μm optical fibers, as indicated, immersed in a transparent homogeneous medium ($n = 1.335$). The horizontal dashed lines represent $z_0 = \frac{\text{core diameter}}{2 \cdot \tan(\text{NA/refractive index})}$

**C** and **D,** Ray tracing simulations of collection efficiency diagrams from a point source ($\lambda = 520$nm) for same fibers as panel (B) immersed in a transparent homogeneous medium ($n = 1.335$) (**C**) or in a turbid medium (Henyey-Greenstein scattering, $n = 1.360$, $l = 48.95$μm, $g = 0.9254$, $T = 0.9989$) (**D**).

**E,** Comparison of axial collection efficiency ($x = 0$, $y = 0$) for 0.22NA/50μm, 0.39NA/200μm, and 0.50NA/200μm optical fibers (panels 1-3 respectively) at $\lambda = 520$nm immersed in a homogeneous medium (blue curve and orange curve for analytical and numerical data, respectively) and in a turbid medium (yellow curve). The red arrows indicate the effect of light collection through the cladding. The horizontal dashed lines represent $z_0$.

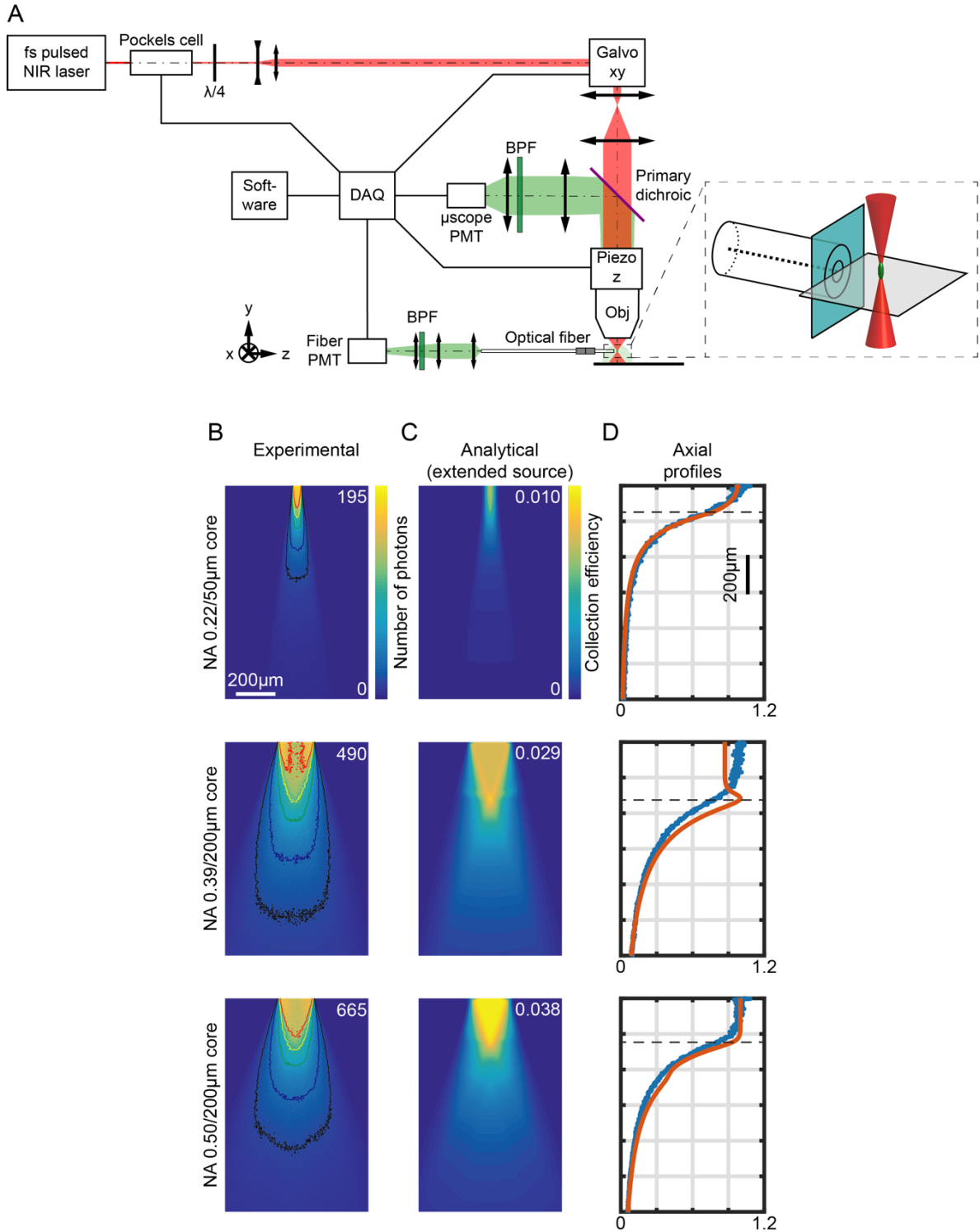

**Figure 2. Measurements of light collection efficiency using 2-photon generated fluorescent point sources**

**A,** Schematic representation of the two-photon microscope used to measure the collection field of optical fibers in quasi-transparent fluorescent medium. The inset shows a magnification of the fiber facet surroundings.

**B,** Section $y = 0$ of the collection field of 0.22/50μm, 0.39/200μm, and 0.50/200μm optical fibers, as indicated, in a 30μM PBS:fluorescein solution, obtained through the *fiber PMT* as shown in panel (A). Isolines at 10%, 20%, 40%, 60%, and 80% of the maximum number mof photons are shown (in black, blue, green, yellow and red, respectively).

**C,** Analytical calculations of collection efficiency diagrams for the three fibers in panel (**B**) immersed in a transparent homogeneous medium ($n = 1.335$) assuming a gaussian source with lateral FWHM $r_{x,z} = 3$μm, axial FWHM $r_y = 32$μm.

**D,** Comparison of normalized experimentally-measured (blue) and analytically-calculated (red) axial collection efficiency profiles ($x = 0$, $y = 0$). The horizontal dashed lines represent $z_0$.

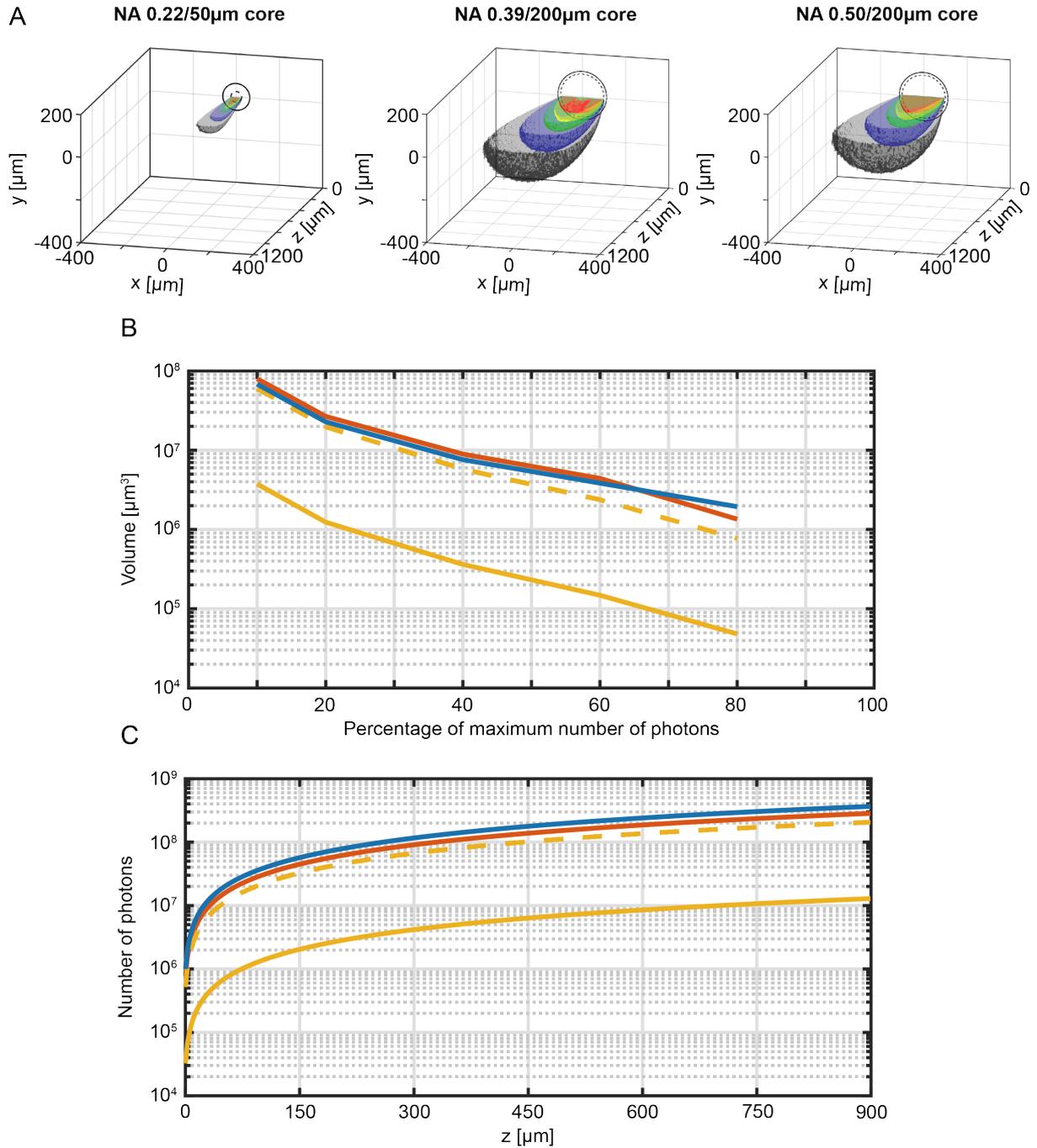

**Figure 3. Calculation of effective light collection volumes**

**A,** Cross-sectional views of the 3-dimensional reconstructions of the collection field for 0.22/50μm, 0.39/200μm, and 0.50/200μm fibers. Iso-intensity surfaces defining the boundaries at which photon collection efficiency falls to 10%, 20%, 40%, 60%, and 80% of its maximum are shown (in black, blue, green, yellow and red, respectively). The continuous and dashed circles in the *xy* plane represent the cladding and the core boundaries, respectively.

**B,** Volumes enclosed by the iso-intensity surfaces shown in panel (A) for 0.22/50μm, 0.39/200μm, and 0.50/200μm fibers (yellow, orange, and blue curves, respectively); the dashed yellow curve represents the data for the 0.22/50μm fiber multiplied by a factor 16 to adjust for the smaller cross-sectional area of this fiber.

**C,** Cumulative number of photons collected by the three fibers as a function of the distance from the fiber facet. Number of photons are shown in a volume 900μm × 600μm × *z*. The dashed yellow curve represents the datapoints relative to the 0.22/50μm fiber multiplied by a factor 16. Error bars are within the line thickness.

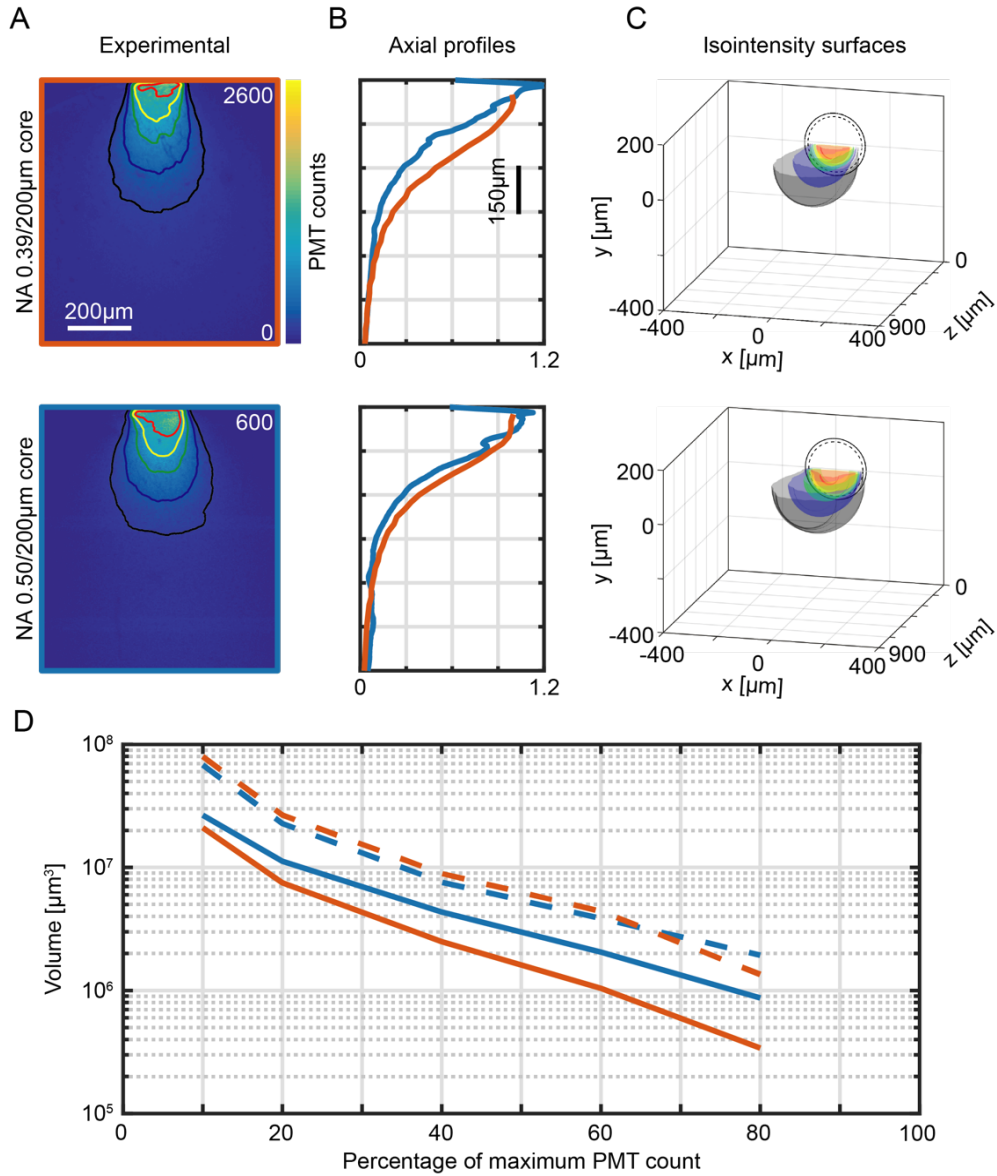

**Figure 4. Photon collection efficiency and effective collective volumes in brain slices**

**A,** Section $y = 0$ of the collection field of 0.39/200μm and 0.50/200μm optical fibers, as indicated, measured in a 300μm thick fluorescently stained brain slice using 2-photon scanning system shown in Figure 2. Isolines at 10%, 20%, 40%, 60%, and 80% of the maximum efficiency are shown (in black, blue, green, yellow and red, respectively).

**B,** Comparison of normalized axial profiles ($x = 0$, $y = 0$) for experimental (in brain slices, blue curves) and numerical data (in turbid medium, red curves) for 0.39/200μm and 0.50/200μm optical fibers.

**C,** Cross-sections of the 3-dimensional reconstruction of the collection field of 0.39/200μm and 0.50/200μm fibers. Iso-intensity surfaces defining the boundaries at which photon collection efficiency falls to 10%, 20%, 40%, 60%, and 80% of its maximum are shown (in black, blue, green, yellow and red, respectively). The continuous and dashed circles represent the cladding and the core boundaries, respectively.

**D,** Volumes enclosed by the iso-intensity surfaces shown in panel (C) for 0.39/200μm and 0.50/200μm fibers (continuous blue and red curves, respectively) compared with volumes enclosed by the iso-intensity surfaces obtained in fluorescein (dashed lines).

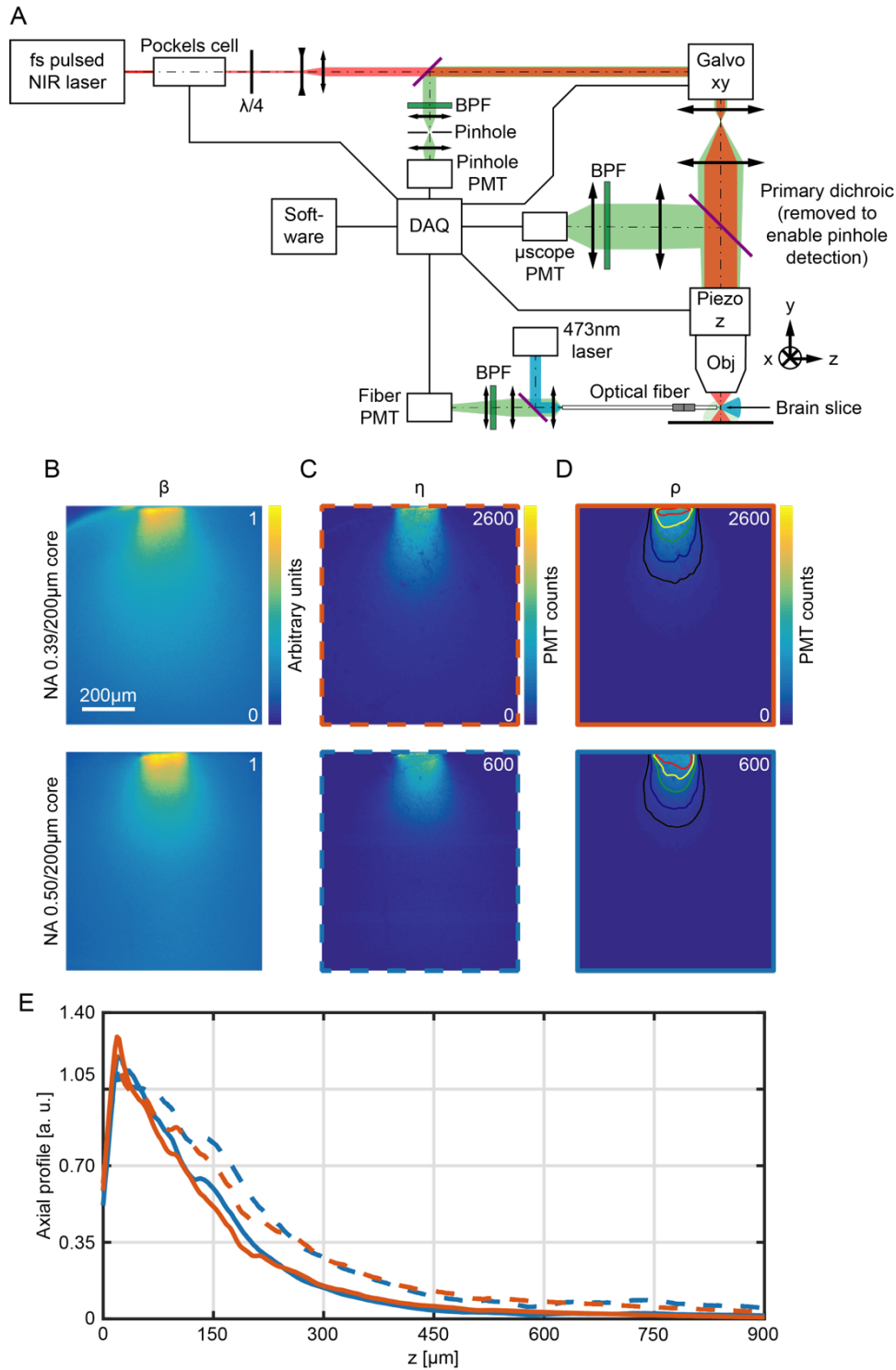

**Figure 5. Scanning pinhole detection of the excitation light field for fiber optics**

**A,** Schematic representation of the optical path used to measure the photometry efficiency diagram of optical fibers in fluorescently stained brain slices.

**B, C and D**, Section $y = 0$ of the normalized light emission diagram $\beta$ (**B**), the collection efficiency $\eta$ (**C**) and the photometry efficiency $\rho$ (**D**) of 0.39/200μm and 0.50/200μm optical fibers, as indicated, measured in a 300μm thick fluorescently stained brain slice. In (D) isolines at 10%, 20%, 40%, 60%, and 80% of the maximum efficiency are shown (in black, blue, green, yellow and red, respectively).

**E,** Comparison of normalized photometry efficiency axial profiles ($x = 0$, $y = 0$) between 0.39/200μm and 0.50/200μm optical fibers (orange and blue continuous curve, respectively). Experimental data from Fig. 4(b1, b2) are reported as comparison (dashed lines).

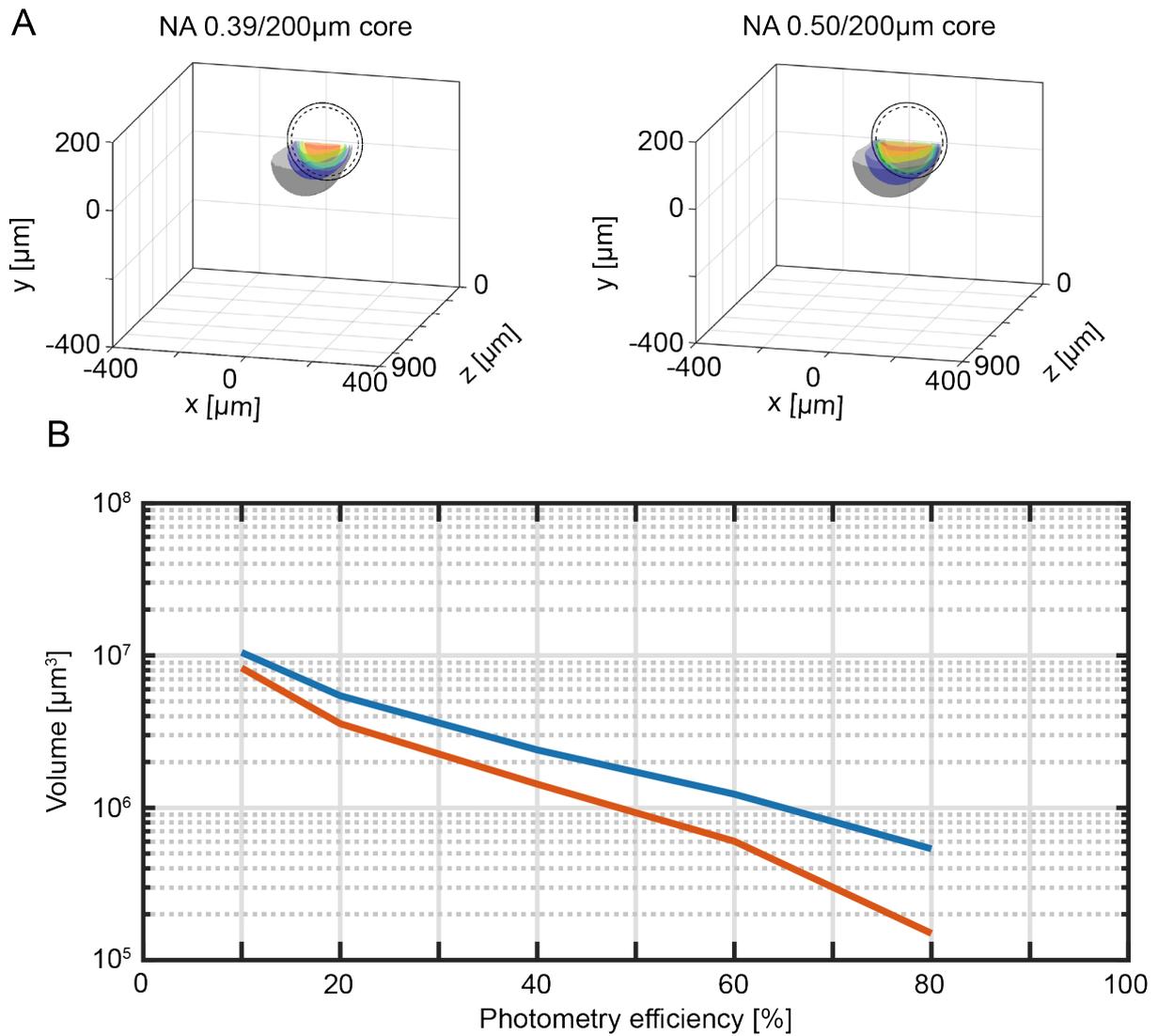

**Figure 6. Effective fiber photometry sampling volumes in tissue**

**A,** Cross-sections of the 3-dimensional reconstruction of the photometry efficiency diagram of 0.39/200µm and 0.50/200µm fibers (panel 1 and 2, respectively). Iso-intensity surfaces at 10%, 20%, 40%, 60%, and 80% efficiency are shown (in black, blue, green, yellow and red, respectively). The continuous and dashed circles represent the cladding and the core boundaries, respectively.

**B,** Volumes enclosed by the iso-intensity surfaces shown in panel (a) for 0.39/200µm and 0.50/200µm fibers (orange and blue curves, respectively).

# Analytical and empirical measurement of fiber photometry signal volume in brain tissue

Marco Pisanello[1, +], Filippo Pisano[1, +], Minsuk Hyun[2, +], Emanuela Maglie[1, 3], Antonio Balena[1, 3], Massimo De Vittorio[1, 3], Bernardo L. Sabatini[2, †, *], Ferruccio Pisanello[1, †, *]

1. Istituto Italiano di Tecnologia, Center for Biomolecular Nanotechnologies, 73010 Arnesano (LE), Italy.
2. Department of Neurobiology, Howard Hughes Medical Institute, Harvard Medical School, 02115 Boston (MA), U.S.A.
3. Dipartimento di Ingegneria dell'Innovazione, Università del Salento, 73100 Lecce (LE), Italy.

+ These authors equally contributed to this work.
† These authors equally contributed to this work.
* Corresponding authors: bernardo_sabatini@hms.harvard.edu - ferruccio.pisanello@iit.it

# Supplementary figures

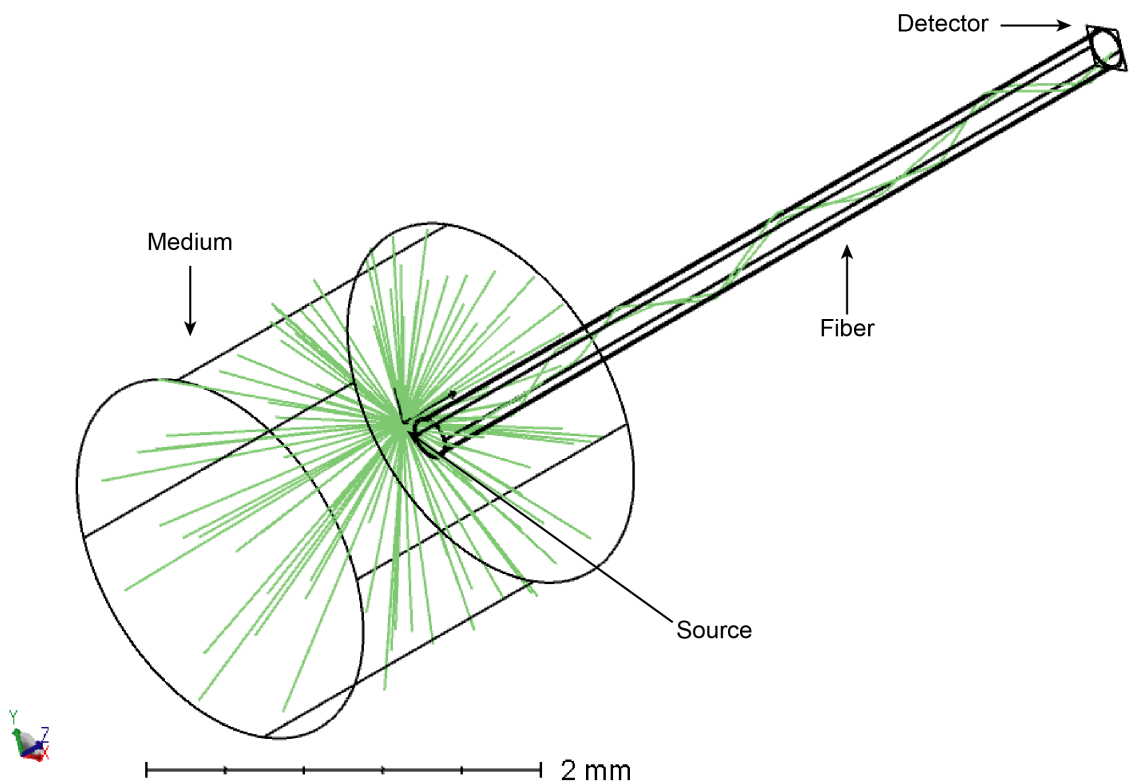

**Supplementary Figure S1.** Ray tracing layout designed to numerically evaluate light collection efficiency of optical fibers.

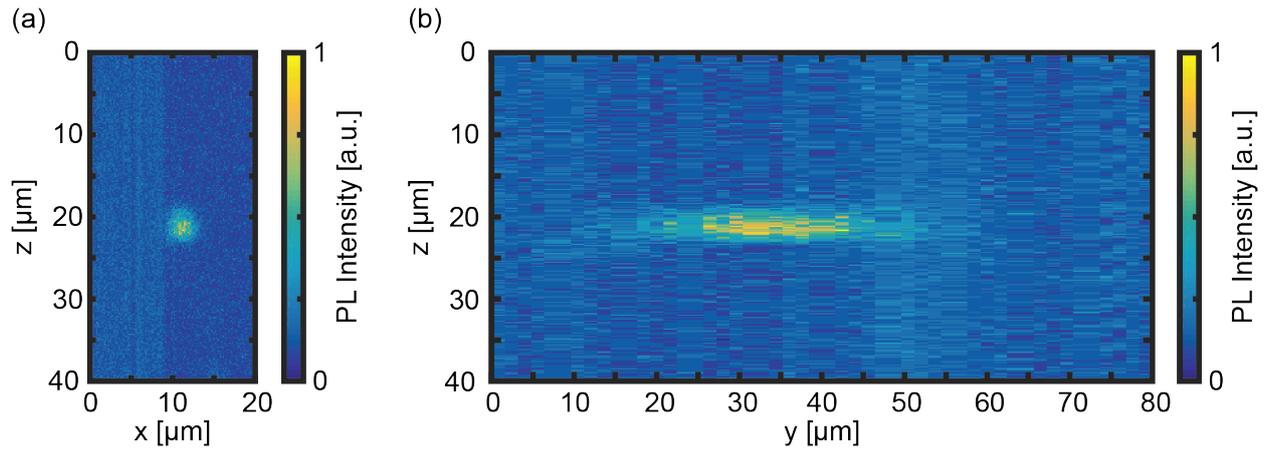

**Supplementary Figure S2. (a):** Lateral section of the custom two-photon microscope point spread function. **(b):** Axial section of the custom two-photon microscope point spread function.

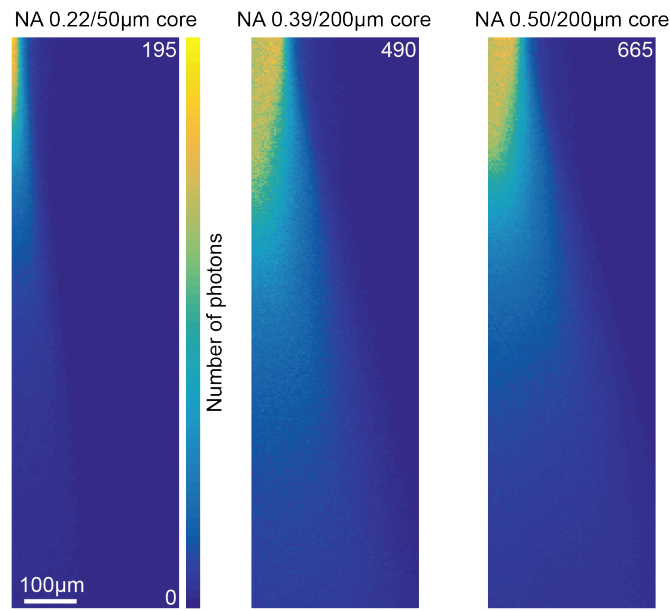

**Supplementary Figure S3.** Section $x = 0$ of the collection diagram of 0.22/50μm, 0.39/200μm, and 0.50/200μm optical fibers, as indicated, in a 30μM PBS:fluorescein solution, obtained through the *fiber PMT* as shown in Fig. 2A.

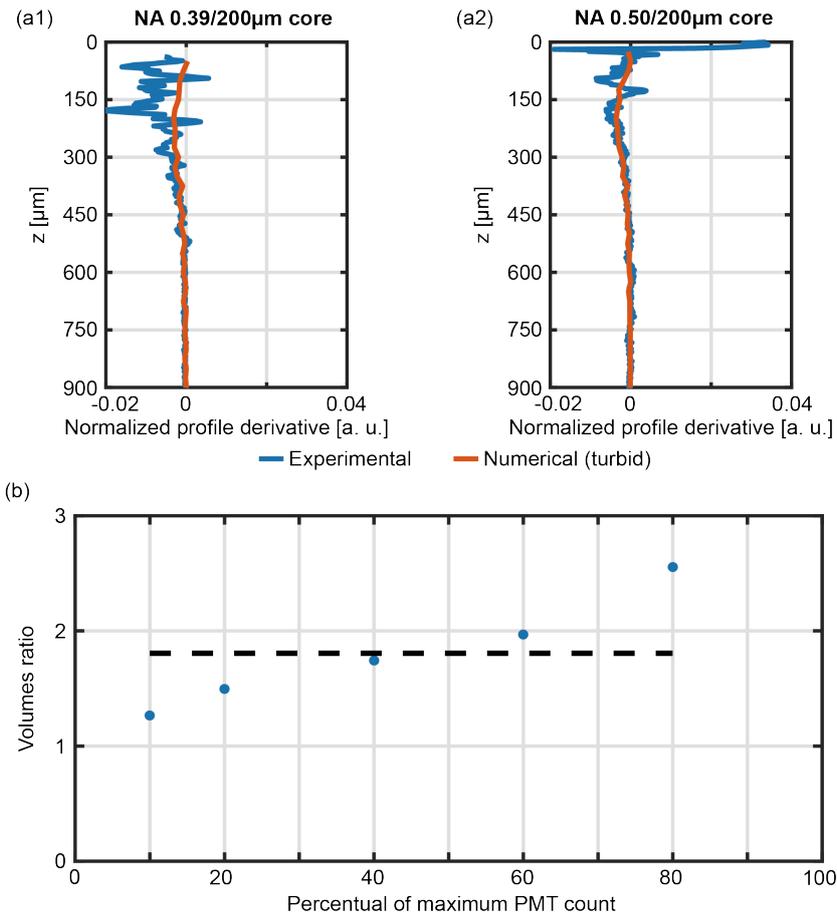

**Supplementary Figure S4. (a):** Comparison of the normalized axial profiles derivatives ($x = 0$, $y = 0$) between experimental and numerical data for 0.39/200μm and 0.50/200μm optical fibers (panel 1 and 2, respectively) in turbid media. **(b):** Ratio of the volumes enclosed by the iso-intensity surfaces shown in Fig. 4(c) (0.50/200μm fiber over 0.39/200μm fiber); the dashed line represents the average of the datapoints.

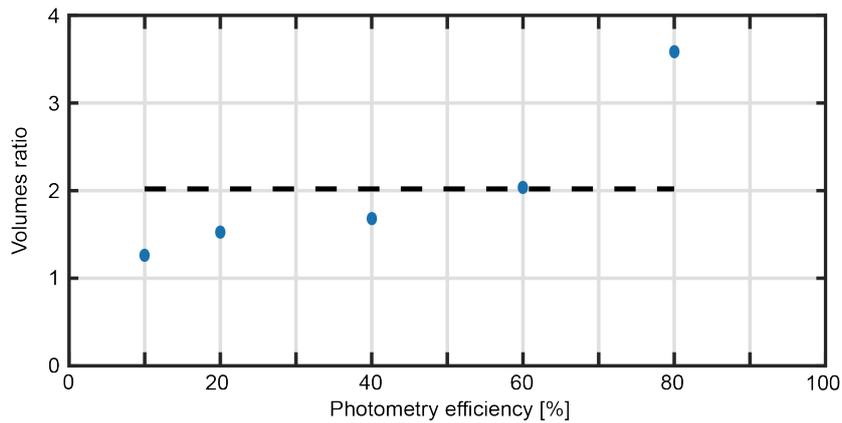

**Supplementary Figure S5.** Ratio of the volumes enclosed by the iso-intensity surfaces shown in Fig. 6(b) (0.50/200μm fiber over 0.39/200μm fiber); the dashed line represents the average of the datapoints.

# Analytical and empirical measurement of fiber photometry signal volume in brain tissue

Marco Pisanello[1, +], Filippo Pisano[1, +], Minsuk Hyun[2, +], Emanuela Maglie[1, 3], Antonio Balena[1, 3], Massimo De Vittorio[1, 3], Bernardo L. Sabatini[2, †, *], Ferruccio Pisanello[1, †, *]

1. Istituto Italiano di Tecnologia, Center for Biomolecular Nanotechnologies, 73010 Arnesano (LE), Italy.
2. Department of Neurobiology, Howard Hughes Medical Institute, Harvard Medical School, 02115 Boston (MA), U.S.A.
3. Dipartimento di Ingegneria dell'Innovazione, Università del Salento, 73100 Lecce (LE), Italy.

+ These authors equally contributed to this work.
† These authors equally contributed to this work.
* Corresponding authors: bernardo_sabatini@hms.harvard.edu - ferruccio.pisanello@iit.it

## Supplementary script

Supplementary script 1: analytical evaluation of collection fields

```matlab
function eta=maps_eta_halfplane(NA,a,n,r,z,visualize)
%maps_eta_halfplane(NA,a,n,r,z) calculate the fiber collection efficiency according
%to Engelbrecht et al. 2009 (doi: 10.1364/OE.17.006421).
%INPUT PARAMETERS
%NA: fiber numerical aperture.
%a: fiber core diameter (in micrometers).
%n: medium refractive index.
%r: radial extension of calculation domain (in micrometers).
%z: axial extension of calculation domain (in micrometers).
%visualize: if equal to 1, the map is plotted.
%OUTPUT VARIABLES
%eta: 2D map of collection efficiency.

    [rg,zg]=meshgrid(r,z);
    a=a/2;

    %Calculates the acceptance cone angle
    alfa=asin(NA/n);
    %Calculates z0
    z0=a/tan(alfa);

    %Calculates the solid angle of the acceptance cone
    omegaNA=2*pi*(1-cos(alfa));
    %Calculates the solid angle of the cone with vertex in an arbitrary
    %position and base on the core facet
    omegaf=2*pi*(1-cos(atan(a./zg.*cos(atan(rg./zg)).^(3/2))));
    %Calculates the solid angle in which emitted power can be accepted by
    %the fiber
    omega=min(omegaf,omegaNA*ones(size(zg)));

    %Normalized coordinates
    u=r/a;
    v=z/z0;
    [ug,vg]=meshgrid(u,v);

    %Calculates the fraction of useful optical fiber reaching the core
```

```matlab
    %facet
    P1=acos((ug.^2-vg.^2+1)./(2*ug));
    P2=vg.^2.*acos((ug.^2+vg.^2-1)./(2*vg.*ug));
    P3=-1/2*sqrt((1-(vg-ug).^2).*((vg+ug).^2-1));
    af=real(P1+P2+P3)./(pi*min(vg,ones(size(vg))).^2);
    af(ug>vg+1)=0;
    af(ug<=max(vg,ones(size(vg)))-min(vg,ones(size(vg))))=1;

    %Calculates collection efficiency
    eta=1/(4*pi)*omega.*af;

    %Visualization of the results.
    if visualize==1
        figure(1);
        imagesc(r,z,eta);
        set(gca,'FontSize',20,'FontWeight','bold','LineWidth',2.5);
        colorbar('LineWidth',2.5);
        xlabel('r [\mum]','FontWeight','bold');
        ylabel('z [\mum]','FontWeight','bold');
        title(strcat('Collection efficiency of',{'
 '},num2str(NA),'NA/',num2str(2*a),'\mum core fibers in n=',num2str(n),'
medium'));
        axis equal;
        axis([min(r) max(r) min(z) max(z)]);

        display(strcat('Collection efficiency of',{'
 '},num2str(NA),'NA/',num2str(2*a),'\mum core fibers in n=',num2str(n),'
medium'));
        display(strcat('Maximum efficiency:',{' '},num2str(max(max(eta)))));
        display(strcat('NA regime distance:',{' '},num2str(z0),'um'));
    end

end

function
[eta_pointsource,eta_convolved]=volume_eta(NA,core_diameter,cladding_diameter,nc
ore,n,r_eta,z_eta,source)
%volume_eta evaluates the analytical 3D collection efficiency diagram for
%both point-like and extended source in quasi transparent medium.
%INPUT PARAMETERS
%NA: fiber numerical aperture.
%core_diameter: fiber core diameter.
%cladding_diameter: fiber cladding external diameter.
%ncore: refractive index of the core.
%n: refractive index of the external medium.
%r_eta: vector of domain points in the lateral direction.
%z_eta: vector of domain points in the axial direction.
%source: matrix modeling the extended source.
%OUTPUT VARIABLES
%eta_pointsource: stack of the collection efficiency with poin-like source.
%eta_convolved: stack of the collection efficiency with extended source.

    %% Generation of the simmetry half-planes (implements eq. 3)
    nclad=sqrt(ncore^2-NA^2);
    NAeq=sqrt(nclad^2-n^2);
    eta_core=maps_eta_halfplane(NA,core_diameter,n,r_eta,z_eta,0);
    eta_clad=maps_eta_halfplane(NAeq,cladding_diameter,n,r_eta,z_eta,0)-
maps_eta_halfplane(NAeq,core_diameter,n,r_eta,z_eta,0);
    eta_halfplane=eta_core+eta_clad;

    %% Volume reconstruction by rotation of the simmetry plane
    [Z,R]=size(eta_halfplane);
    eta_pointsource=zeros(2*R-1,2*R-1,Z);
```

```matlab
    [i1,i2,i3]=ndgrid(1:2*R-1,1:2*R-1,1:Z);

    x=i1-R;
    y=i2-R;
    z=i3-1;

    [~,r_out,z_out]=cart2pol(x,y,z);

    [j1,j2]=meshgrid(1:R,1:Z);
    r_in=j1-1;
    z_in=j2-1;

eta_pointsource(:)=griddata(r_in,z_in,eta_halfplane,r_out(:),z_out(:),'linear');
    eta_pointsource(isnan(eta_pointsource))=0;

    %% Convolution between source and collection efficiency
    eta_convolved=convn(eta_pointsource,source,'same');
    eta_convolved(isnan(eta_convolved))=0;
end

function source=source(lateral,axial,domain)
% source models the microscope PSF as a gaussian fluorescent spot
% INPUT PARAMETERS
% lateral: FWHM of the PSF in the lateral dimensions [um]
% axial: FWHM of the PSF in the axial dimension [um]
% domain: axis of the domain in which the source should be represented [um]
% OUTPUT VARIABLES
% source: representation of the PSF
    sigmasource_x=lateral/(2*sqrt(2*log(2)));
    sigmasource_y=axial/(2*sqrt(2*log(2)));
    sigmasource_z=lateral/(2*sqrt(2*log(2)));
    [xmat_source,ymat_source,zmat_source]=meshgrid(domain,domain,domain);
    source=exp(-
(xmat_source.^2/(2*sigmasource_x^2)+ymat_source.^2/(2*sigmasource_y^2)+zmat_sour
ce.^2/(2*sigmasource_z^2)));
    source=source/(sum(reshape(source,[],1)));
end
```

Supplementary script 2: data processing to obtain collection fields in quasi-transparent medium

```matlab
%% Processing of stack acquired in fluorescein

%% User-defined variables

% Fiber parameters
NA=0.50; %Numerical aperture
core_diameter=200; %Core diameter [?m]
cladding_diameter=225; %Cladding diameter [?m]
ncore=1.4613; %Core refractive index

% Medium parameters
n=1.335; %Medium refractive index

% Dataset files
filename_gain='gain_00001_00001.tif'; %Image for gain measurement
filename_stack='stack_00001_00001.tif'; %Volumetric stack

% Measurement parameters
z_focus=6; %Slice # at which the focal plane contains the optical axis
x_step=2.7; %Pixel size in the lateral direction [um]
z_step=5; %Pixel size in the axial direction [um]
```

```matlab
rotation=0; %Rotation of the image [degree]
axis_index=244; %Pixel # individuating the optical axis
core_index=42; %Pixel # individuating the core facet
discard=0; %Number of pixel to discard due to rotation (zero padding)
n_of_slice=80; %Number of slice in the stack
n_of_frame=5; %Number of frame acquired for each slice
image_size=512; %Number of pixel of each slice side
photon_top_022=109.7259; %Normalization parameters for fluorescence excitation

% Analytical calculation parameters
r_an_step=5; %Step size along the lateral direction [?m]
z_an_step=5; %Step size along the axial direction [?m]
r_an_max=500; %Domain size along the lateral direction [?m]
z_an_max=1200; %Domain size along the lateral direction [?m]
r_an=[0:r_an_step:r_an_max];
z_an=[0:z_an_step:z_an_max];
r_an_full=horzcat(-fliplr(r_an(2:end)),r_an);

% Simulation parameters
grid_step_rt=25; %Step size of the ray tracing grid [?m]
r_rt=[0:19]*grid_step_rt;
z_rt=[0:60]*grid_step_rt;

% Source definition
source_lateral=3; %Lateral extension of the microscope PSF [?m]
source_axial=32; %Axial extension of the microscope PSF [?m]
source_domain=[-50:5:50]; %Side of the domain for source representation [?m]

% Processing parameters
smooth_factor_3d=3; %Size of the smoothing applied to the stack
smooth_factor_2d=3; %Size of the smoothing applied to the ray tracing map
smooth3d_enable=0; %If equal to 1, smoothing of the stack is enabled
smooth2d_enable=0; %If equal to 1, smoothing of the ray tracing is enabled
distance_mean_profile=30; %Number of pixel to average for profile normalization

% Plots parameters
longitudinal_step=50; %Step for the longitudinal planes visualization [?m]
transversal_step=100; %Step for the transversal planes visualization [?m]

%% END OF USER-DEFINED SECTION
% Do not edit below unless you really want to.

%% Load data (if not already in the workspace)
if ~exist('bench','var')
    display('Loading data');
    [gain_top,gain_bench,photon_top]=read_gain(filename_gain); %Calculates the gain of both acquisition channels and acquires the mean number of photon acquired from the microscope PMT

[top,bench,z,x,y]=read_stack(filename_stack,n_of_slice,n_of_frame,image_size,image_size,z_step,x_step,x_step,1,1); %Reads the stack acquired from both channels
    top(top>max(reshape(top(1:512-core_index,:,:),1,[])))=0; %Removes outliers in the volume occupied by the fiber
    import_rt; %Reads data from ray tracing simulations in transparent medium
    import_rtb; %Reads data from ray tracing simulations in turbid medium
    RT(:,3)=[]; %Removes unnecessary data
    RT_brain(:,3)=[]; %Removes unnecessary data
end

%% Pre-processing of data: baseline removal, smoothing, and rotation
display('Data pre-processing');
bench_positive=bench-min(reshape(bench,1,[])); %Removes baseline
bench_smooth=bench_positive;
```

```matlab
top_smooth=top;
if(smooth3d_enable==1) %Smoothing of data (if enabled)
    bench_smooth=smooth3(bench_positive,'box',smooth_factor_3d);
    top_smooth=smooth3(top,'box',smooth_factor_3d);
end
bench_smooth=flip(imrotate(bench_smooth,rotation,'bilinear'),1); %Rotates images
top_smooth=flip(imrotate(top_smooth,rotation,'bilinear'),1);

%% PMT count to photon number conversion and flat-field correction
top_normalized=normalize(top_smooth); %Normalizes stack acquired from the top PMT
bench_photon=(photon_top_022/photon_top)*bench_smooth./(gain_bench*top_normalized); %Converts the stack acquired from the fiber PMT in number of photons
bench_photon(bench_photon==Inf)=0; %Removes outliers

%% Errors determination
sigma_r_bench=sqrt(bench_smooth)./bench_smooth;
sigma_r_top=sqrt(top_smooth)./top_smooth;
sigma=sqrt(sigma_r_bench.^2+sigma_r_top.^2).*bench_photon;

%% Definition of the central slice for data analysis of measurement
eta_measure=bench_photon(:,:,z_focus);

%% Definition of the axial profile and integral on measurement
eta_profile=eta_measure(core_index:end-discard,axis_index);

%% Definition of analytical data (if not already in the workspace)
if ~exist('eta_pointsource','var')
    display('Computation of analytical data');
    nclad=sqrt(ncore^2-NA^2); %Evaluates cladding refractive index
    NAeq=sqrt(nclad^2-n^2); %Evaluate equivalent numerical aperture at the interface cladding/medium
    source=source(source_lateral,source_axial,source_domain); %Defines the fluorescent spot

[eta_pointsource,eta_convolved]=volume_eta(NA,core_diameter,cladding_diameter,ncore,n,r_an,z_an,source); %Evaluates the analytical collection efficiency
    eta_pointsource=permute(eta_pointsource,[3 1 2]);
    eta_convolved=permute(eta_convolved,[3 1 2]);
    source=permute(source,[3 1 2]);
end

%% Definition of numerical data
display('Processing of numerical data');
eta_rt=RT_to_eta(RT,length(z_rt),grid_step_rt,grid_step_rt,0);
eta_rt_brain=RT_to_eta(RT_brain,length(z_rt),grid_step_rt,grid_step_rt,0);
eta_rt=flipud(eta_rt);
eta_rt_brain=flipud(eta_rt_brain);
if(smooth2d_enable==1) %Smoothing of data (if enabled)

eta_rt=filter2((1/smooth_factor_2d)^2*ones(smooth_factor_2d,smooth_factor_2d),eta_rt);
end

%% Evaluation of normalized collection volumes (if not already in the workspace)
volumes=[10 20 40 60 80]; %Percentages of maximum at which volumes should be evaluated
if ~exist('volumes_measurement','var')
    display('Computation of measured and analytical collection volumes');
    temp=cat(3,bench_photon(:,:,end:-1:z_focus),bench_photon(:,:,z_focus+1:end)); %Concatenates the stack with a mirrored version of itself to obtain a full volume
    [xx,~,zz]=size(temp);
```

```matlab
    [~,volumes_measurement]=eval_isosurfaces([0:xx-1]*x_step,[0:xx-1]*x_step,[0:zz-1]*z_step,temp,3); %
    clear xx zz temp;
end

%% Evaluation of cumulative number of photons (if not already in the workspace)
if ~exist('cumulative_photons','var')
    temp=cat(3,bench_photon(:,:,end:-1:z_focus+1),bench_photon(:,:,z_focus:end)); %Concatenates the stack with a mirrored version of itself to obtain a full volume
    temp2=cat(3,sigma(:,:,end:-1:z_focus+1),sigma(:,:,z_focus:end));
    [~,~,z_center]=size(bench_photon(:,:,z_focus:end));
    
[cumulative_photons,plane_photons]=count_photons_in_volume(x_step,z_step,core_index,axis_index,z_center,temp,temp2);
    clear temp temp2;
end

%% Plots
% Longitudinal planes (y=const)
figure(1);
for i=1:6
    subplot(3,6,i);
    imagesc(r_an_full,z_an,squeeze(eta_pointsource(:,:,length(r_an)+(i-1)*round(longitudinal_step/z_an_step))),[0 max(reshape(eta_pointsource,[],1))]);
    axis equal;
    axis([-max(r_an) max(r_an) min(z_an) max(z_an)]);
    title(strcat('y=',num2str(r_an_full(length(r_an)+(i-1)*round(longitudinal_step/z_an_step))),'\mum'));
    subplot(3,6,6+i);
    imagesc(r_an_full,z_an,squeeze(eta_convolved(:,:,length(r_an)+(i-1)*round(longitudinal_step/z_an_step))),[0 max(reshape(eta_convolved,[],1))]);
    axis equal;
    axis([-max(r_an) max(r_an) min(z_an) max(z_an)]);
    title(strcat('y=',num2str(r_an_full(length(r_an)+(i-1)*round(longitudinal_step/z_an_step))),'\mum'));
    subplot(3,6,12+i);
    temp=bench_photon(:,:,z_focus+(i-1)*round(longitudinal_step/z_step));
    [xx,~]=size(temp);
    imagesc(([0:xx-1]-axis_index)*x_step,([0:xx-1]-core_index)*x_step,temp,[0 max(reshape(bench_photon(:,:,z_focus),[],1))]);
    axis equal;
    axis([-max(r_an) max(r_an) min(z_an) max(z_an)]);
    title(strcat('y=',num2str(z(z_focus+(i-1)*round(longitudinal_step/z_step))-z(z_focus)),'\mum'));
end
savefig('Figures/longitudinal_planes.fig');

% Transversal planes (z=const)
figure(2);
for i=1:6
    subplot(3,6,i);
    imagesc(r_an_full,r_an_full,squeeze(eta_pointsource(1+(i-1)*round(transversal_step/r_an_step),:,:)),[0 max(reshape(eta_pointsource,[],1))]);
    axis equal;
    axis([-max(r_an) max(r_an) -max(r_an) max(r_an)]);
    title(strcat('z=',num2str(z_an(1+(i-1)*round(transversal_step/r_an_step))),'\mum'));
    subplot(3,6,6+i);
    imagesc(r_an_full,r_an_full,squeeze(eta_convolved(1+(i-1)*round(transversal_step/r_an_step),:,:)),[0 max(reshape(eta_convolved,[],1))]);
    axis equal;
```

```matlab
        axis([-max(r_an) max(r_an) -max(r_an) max(r_an)]);
        title(strcat('z=',num2str(z_an(1+(i-
1)*round(transversal_step/r_an_step))),'\mum'));
        subplot(3,6,12+i);
        temp=flipud(squeeze(bench_photon(core_index+(i-
1)*round(transversal_step/x_step),:,z_focus:end))');
        [z_center,~]=size(temp);
        temp=vertcat(temp,flipud(temp(2:end,:)));
        [zz,xx]=size(temp);
        imagesc(([0:xx-1]-axis_index)*x_step,([0:zz-1]-z_center)*z_step,temp,[0
max(reshape(bench_photon(:,:,z_focus),[],1))]);
        axis equal;
        axis([-max(r_an) max(r_an) -max(r_an) max(r_an)]);
        title(strcat('z=',num2str(abs(x(core_index+(i-
1)*round(transversal_step/x_step))-x(core_index))),'\mum'));
        clear temp;
end
savefig('Figures/transversal_planes.fig');

% Plots normalized axial profiles
index_mean_profile=round(distance_mean_profile/x_step);
figure(3);
set(gca,'FontSize',20,'FontWeight','bold','LineWidth',2.5);
xlabel('Distance from the facet [\mum]','FontWeight','bold');
ylabel('Normalized axial profile [a.u.]','FontWeight','bold');
title(strcat('Axial profiles comparison for a',{'
'},num2str(NA),'NA/',num2str(core_diameter),'\mum core fibers in fluorescein'));
hold on;
plot([0:length(eta_profile)-
1]*x_step,eta_profile/mean(eta_profile(1:index_mean_profile)),'LineWidth',3);
plot(z_rt(2:end),eta_rt(2:end,length(r_rt))/max(eta_rt(:,length(r_rt))),'LineWid
th',3);
plot(z_an,squeeze(eta_pointsource(:,length(r_an),length(r_an)))/max(eta_pointsou
rce(:,length(r_an),length(r_an))),'LineWidth',3,'Color',[0.93 0.69 0.13]);
plot(z_an,squeeze(eta_convolved(:,length(r_an),length(r_an)))/max(eta_convolved(
:,length(r_an),length(r_an))),'LineWidth',3,'Color',[0.49 0.18 0.56]);
legend('Experimental','Numerical','Analytical (point source)','Analytical
(extended source)');
box();
grid();
axis([min(z_an) max(z_an) 0 1.2]);
hold off;
savefig('Figures/axial_profiles.fig');

% Plots normalized collection volumes
figure(4);
set(gca,'FontSize',20,'FontWeight','bold','LineWidth',2.5,'YScale','log');
xlabel('% of maximum number of photons','FontWeight','bold');
ylabel('Volume [\mum^3]','FontWeight','bold');
title(strcat('Collection volumes for a',{'
'},num2str(NA),'NA/',num2str(core_diameter),'\mum core fibers in fluorescein'));
hold on;
plot(volumes,volumes_measurement,'LineWidth',3);
legend('Experimental','Analytical (point source)','Analytical (extended
source)');
box();
grid();
axis([0 100 1e5 1e9]);
hold off;
savefig('Figures/normalized_volumes.fig');

% Plots cumulative number of photons
figure(5);
set(gca,'FontSize',20,'FontWeight','bold','LineWidth',2.5);
```

```matlab
xlabel('z [\mum]','FontWeight','bold');
ylabel('Collected photons per plane (cumulative)','FontWeight','bold');
title(strcat('Photons collected from a',{' '},num2str(NA),'NA/',num2str(core_diameter),'\mum core fibers in fluorescein'));
hold on;
plot([0:length(cumulative_photons)-1]*x_step,cumulative_photons,'LineWidth',3);
box();
grid();
hold off;
savefig('Figures/cumulative_photons.fig');

% Plot efficiency maps
figure(6);
[xx,zz]=size(eta_measure);
imagesc(([0:xx-1]-axis_index)*x_step,([0:zz-1]-core_index)*x_step,eta_measure);
axis equal;
axis([-max(r_an) max(r_an) min(z_an) max(z_an)]);
set(gca,'FontSize',20,'FontWeight','bold','LineWidth',2.5);
title('Experimental collection efficiency map');
xlabel('x [\mum]');
ylabel('z [\mum]');
savefig('Figures/experimental_map.fig');

figure(7);
imagesc(r_an_full,z_an,squeeze(eta_pointsource(:,:,length(r_an))));
axis equal;
axis([-max(r_an) max(r_an) min(z_an) max(z_an)]);
set(gca,'FontSize',20,'FontWeight','bold','LineWidth',2.5);
title('Analytical (point source) collection efficiency map');
xlabel('x [\mum]');
ylabel('z [\mum]');
savefig('Figures/pointsource_map.fig');

figure(8);
imagesc(r_an_full,z_an,squeeze(eta_convolved(:,:,length(r_an))));
axis equal;
axis([-max(r_an) max(r_an) min(z_an) max(z_an)]);
set(gca,'FontSize',20,'FontWeight','bold','LineWidth',2.5);
title('Analytical (extended source) collection efficiency map');
xlabel('x [\mum]');
ylabel('z [\mum]');
savefig('Figures/convolved_map.fig');

figure(9);
r_rt_full=[-fliplr(r_rt(2:end)) r_rt];
imagesc(r_rt_full,z_rt,eta_rt);
axis equal;
axis([-max(r_an) max(r_an) min(z_an) max(z_an)]);
set(gca,'FontSize',20,'FontWeight','bold','LineWidth',2.5);
title('Numerical collection efficiency map');
xlabel('x [\mum]');
ylabel('z [\mum]');
savefig('Figures/numerical_map.fig');

figure(10);
imagesc(r_rt_full,z_rt,eta_rt_brain);
axis equal;
axis([-max(r_an) max(r_an) min(z_an) max(z_an)]);
set(gca,'FontSize',20,'FontWeight','bold','LineWidth',2.5);
title('Numerical collection efficiency map (in brain)');
xlabel('x [\mum]');
ylabel('z [\mum]');
savefig('Figures/numerical_map_brain.fig');
```

```matlab
% Plots axial profiles (analytical and numerical without normalization)
figure(11);
set(gca,'FontSize',20,'FontWeight','bold','LineWidth',2.5);
xlabel('Distance from the facet [\mum]','FontWeight','bold');
ylabel('Normalized axial profile [a.u.]','FontWeight','bold');
title(strcat('Axial profiles comparison for a',{'
'},num2str(NA),'NA/',num2str(core_diameter),'\mum core fibers in fluorescein'));
hold on;
plot(z_an,squeeze(eta_pointsource(:,length(r_an),length(r_an))),'LineWidth',3);
plot(z_rt(2:end),eta_rt(2:end,length(r_rt)),'LineWidth',3);
plot(z_rt(2:end),eta_rt_brain(2:end,length(r_rt)),'LineWidth',3);
legend('Analytical (point source)','Numerical (homogeneous)','Numerical (turbid)');
box();
grid();
axis([min(z_an) max(z_an) 0 0.038]);
hold off;
savefig('Figures/axial_profiles_fig1.fig');
```

Supplementary script 3: data processing to obtain collection fields in brain slice

```matlab
%% Processing of images acquired in brain slice

%% User defined variables
axis_index=263; %Pixel number identifying central slice.
core_index=122; %Pixel number identifying core facet.
x_step=2.7; %Step in the lateral direction.
n_of_frame=60; %Number of frame averaged in the acquisition.
image_size=512; %Image size in pixel (square image is assumed).

volumes=[10 20 40 60 80]; %Percentage at which isosurface are determined.

smooth_size=1; %Smoothing window size for volume determination.

%% Processing
%Reads measurement data
if ~exist('bench','var')

[top,bench,x,y]=read_grab('grab_00001_00001.tif',n_of_frame,image_size,image_size,x_step,x_step,1,1);
    bench=bench-min(reshape(bench,1,[]));
    bench=filter2(fspecial('average',11),bench);
    halfplane=bench';
    halfplane=halfplane(core_index:end,1:axis_index);
    halfplane=fliplr(halfplane);
end

%Reconstruction of 3D data from central slice (measurements)
if ~exist('volume_reconstructed','var')
    temp=volume_reconstruction(halfplane);
    volume_reconstructed=permute(temp,[3 1 2]);
    clear temp;
end

%Volume calculation
temp=cat(3,volume_reconstructed(:,:,end:-1:axis_index),volume_reconstructed(:,:,axis_index+1:end));
[xx,yy,zz]=size(temp);
temp(temp>500)=0;
[~,volumes_measurement]=eval_isosurface([0:xx-1]*x_step,[0:yy-1]*x_step,[0:zz-1]*x_step,temp,smooth_size);
```

```matlab
clear temp xx yy zz;

%Isosurface visualization
temp=volume_reconstructed(:,:,end:-1:axis_index);
[xx,yy,zz]=size(temp);
temp(temp>500)=0;
[~,~]=eval_isosurface([0:xx-1]*x_step,[0:yy-1]*x_step,[0:zz-
1]*x_step,temp,smooth_size,0,1);
clear temp xx yy zz;

%Plot of volumes
figure(2);
plot(volumes,volumes_measurement);
```

Supplementary script 4: data processing to obtain photometry efficiency

```matlab
%% Processing of images acquired in brain slice to determine rho

%% User defined variables
axis_index=290; %Pixel number identifying central slice.
core_index=142;  %Pixel number identifying core facet.
rotazione=4; %Image rotation (in degree)

volumes=[10 20 40 60 80]; %Percentage at which isosurface are determined.

smooth_size=1; %Smoothing window size for volume determination.
x_step=2.7; %Step in the lateral direction.
n_of_frame=60; %Number of frame averaged in the acquisition.
image_size=512; %Image size in pixel (square image is assumed).

%% Processing
%Reads measurement data
if ~exist('bench','var')

[top,bench,x,y]=read_grab('grab_00001_00001.tif',n_of_frame,image_size,image_siz
e,x_step,x_step,1,1);

[pinhole,~,~,~]=read_grab('ph_100_00001_00001.tif',n_of_frame,image_size,image_s
ize,x_step,x_step,1,1);
end

%Registers images
[optimizer,metric]=imregconfig('multimodal');
registered=imregister(pinhole,bench,'translation',optimizer,metric);

%Pixel-by-pixel product
efficiency=imrotate(bench/max(reshape(bench,1,[])).*registered/max(reshape(regis
tered,1,[])),rotazione,'bilinear');
[n_of_pixel,~]=size(efficiency);

%Smoothing of data
if ~exist('halfplane','var')
    halfplane=efficiency';
    halfplane=filter2(fspecial('average',11),halfplane);
    halfplane=halfplane(core_index:end,1:axis_index);
    halfplane=fliplr(halfplane);
end

%Volume reconstruction
if ~exist('volume_reconstructed','var')
    temp=volume_reconstruction(halfplane);
    volume_reconstructed=permute(temp,[3 1 2]);
```

```matlab
        clear temp;
    end

    %Volumes determination and isosurfaces visualization
    if ~exist('volumes_measurement','var')
        temp=cat(3,volume_reconstructed(:,:,end:-
1:axis_index),volume_reconstructed(:,:,axis_index+1:end));
        [xx,yy,zz]=size(temp);
        [~,volumes_measurement]=superficidilivello([0:xx-1]*x_step,[0:yy-
1]*x_step,[0:zz-1]*x_step,temp,smooth_size);
        clear temp xx yy zz;

        temp=volume_reconstructed(:,:,end:-1:axis_index);
        [xx,yy,zz]=size(temp);
        [~,~]=superficidilivello([0:xx-1]*x_step,[0:yy-1]*x_step,[0:zz-
1]*x_step,temp,smooth_size,0,1);
        clear temp xx yy zz;
    end

    %% Plots of results
    %Plots of image acquired through microscope and fiber PMT (2P excitation),
    %through pinhole PMT (1P excitation), and registering of fiber and pinhole
    %PMT channels
    figure(2);
    subplot(2,2,1);
    imagesc(([1:n_of_pixel]-core_index)*2.7,([1:n_of_pixel]-
axis_index)*2.7,imrotate(top,rotazione,'bilinear'));
    axis equal;
    axis([0 900 -400 400]);
    subplot(2,2,2);
    imagesc(([1:n_of_pixel]-core_index)*2.7,([1:n_of_pixel]-
axis_index)*2.7,imrotate(bench,rotazione,'bilinear'));
    axis equal;
    axis([0 900 -400 400]);
    subplot(2,2,3);
    imagesc(([1:n_of_pixel]-core_index)*2.7,([1:n_of_pixel]-
axis_index)*2.7,imrotate(pinhole,rotazione,'bilinear'));
    axis equal;
    axis([0 900 -400 400]);
    subplot(2,2,4);
    imagesc(([1:n_of_pixel]-core_index)*2.7,([1:n_of_pixel]-
axis_index)*2.7,imrotate(registered,rotazione,'bilinear'));
    axis equal;
    axis([0 900 -400 400]);

    %Plots efficiency map
    figure(3);
    imagesc(([1:n_of_pixel]-core_index)*2.7,([1:n_of_pixel]-
axis_index)*2.7,efficiency);
    axis equal;
    axis([0 900 -400 400]);

    %Plots volumes
    figure(4);
    plot(volumes,volumes_measurement);
```

Supplementary script 5: other functions

```matlab
function
[cumulative_photons,plane_photons,sigma_cumulative]=count_photons_in_volume(x_st
ep,z_step,core_index,axis_index,focus_index,stack,sigma)
%count_photons_in_volume Evaluates the cumulative number of photons and its
uncertainty from a
%volumetric stack in a 800x400x900 um^3 volume.
```

```matlab
%INPUT PARAMETERS
%x_step: step size along lateral and axial direction.
%z_step: step size along transversal direction.
%core_index: pixel number identifying core facet.
%core_index: pixel number identifying fiber axis in the lateral direction.
%focus_index: pixel number identifying fiber axis in the transversal direction.
%stack: volumetric stack from the measurement.
%sigma: uncertainty on stack.
%OUTPUT VARIABLES
%cumulative_photons: vector of cumulative number of photons.
%plane_photons: vector of number of photons collected from single planes.
%sigma_cumulative: uncertainty on cumulative_photons.

    n_focus=ceil_even(400/z_step);
    n_axial=round(900/x_step);
    n_lateral=ceil_even(800/x_step);
    
    stack_reduced=stack(core_index:core_index+n_axial,axis_index-
n_lateral/2:axis_index+n_lateral/2,focus_index-n_focus/2:focus_index+n_focus/2);
    sigma_reduced=sigma(core_index:core_index+n_axial,axis_index-
n_lateral/2:axis_index+n_lateral/2,focus_index-n_focus/2:focus_index+n_focus/2);
    plane_photons=squeeze(sum(sum(stack_reduced,3),2));
    
    cumulative_photons=cumsum(plane_photons);
    sigma_plane=squeeze(sum(sum(sigma_reduced.^2,3),2));
    sigma_cumulative=sqrt(cumsum(sigma_plane));

end

function [maximum,volumes]=eval_isosurfaces(x,y,z,stack,smooth_size,varargin)
%eval_isosurfaces traces the isosurfaces on a volumetric stack. Maximum
%value of the stack as well volumes enclosed within the isosurfaces at 10%,
%20%, 40%, 60% and 80% of the maximum are returned.
%INPUT PARAMETERS
%x: distance vector along lateral dimension x.
%y: distance vector along lateral dimension y.
%x: distance vector along axial dimension z.
%stack: volumetric stack.
%smooth_size: smoothing intensity (must be an odd integer)
%OPTIONAL INPUT PARAMETERS
%reverse: if equal to 1 inverts z axis visualization.
%enable_output: if equal to 1 enables the plot of outputs.
%ctrl: if equal to 1 enables the visualization of voxel employed to
%determine enclosed volumes for each isosurfaces
%OUTPUT VARIABLES
%maximum: maximum value in the stack.
%volumes: volumes enclosed within the isosurfaces at 10%,
%20%, 40%, 60% and 80% of the maximum

    if length(varargin)>=1
        reverse=varargin{1};
    end
    if length(varargin)>=2
        enable_output=varargin{2};
    end
    if length(varargin)>=3
        ctrl=varargin{3};
    end
    
    if(~exist('enable_output','var'))
        enable_output=0;
    end
```

```matlab
    stack=smooth3(stack,'box',smooth_size);
    maximum=max(max(max(stack)));
    if(enable_output==1)
        display('Maximum number of photon: ');
        display(maximum);
    end

    [X,Y,Z]=ndgrid(x,y,z);
    dx=abs(x(2)-x(1));
    dy=abs(y(2)-y(1));
    dz=abs(z(2)-z(1));

    s10=isosurface(X,Y,Z,stack,0.1*maximum);
    vox10=double(stack>=0.1*maximum);
    v10=sum(sum(sum(vox10)))*dx^2*dz;
    if(enable_output==1)
        display(['Volume at 10% of maximum (' num2str(maximum*0.1) ') [um^3]: ']);
        display(v10);
    end

    s20=isosurface(X,Y,Z,stack,0.2*maximum);
    vox20=double(stack>=0.2*maximum);
    v20=sum(sum(sum(vox20)))*dx^2*dz;
    if(enable_output==1)
        display(['Volume at 20% of maximum (' num2str(maximum*0.2) ') [um^3]: ']);
        display(v20);
    end

    s40=isosurface(X,Y,Z,stack,0.4*maximum);
    vox40=double(stack>=0.4*maximum);
    v40=sum(sum(sum(vox40)))*dx^2*dz;
    if(enable_output==1)
        display(['Volume at 40% of maximum (' num2str(maximum*0.4) ') [um^3]: ']);
        display(v40);
    end

    s60=isosurface(X,Y,Z,stack,0.6*maximum);
    vox60=double(stack>=0.6*maximum);
    v60=sum(sum(sum(vox60)))*dx^2*dz;
    if(enable_output==1)
        display(['Volume at 60% of maximum (' num2str(maximum*0.6) ') [um^3]: ']);
        display(v60);
    end

    s80=isosurface(X,Y,Z,stack,0.8*maximum);
    vox80=double(stack>=0.8*maximum);
    v80=sum(sum(sum(vox80)))*dx^2*dz;
    if(enable_output==1)
        display(['Volume at 80% of maximum (' num2str(maximum*0.8) ') [um^3]: ']);
        display(v80);
    end

    volumes=[v10 v20 v40 v60 v80];

    if(enable_output==1)
        figure(1);
        hold on;
```

```matlab
        p10=patch(s10);
        p10.FaceColor='k';
        p10.FaceAlpha=0.25;
        p10.LineStyle='none';
        
        p20=patch(s20);
        p20.FaceColor='b';
        p20.FaceAlpha=0.25;
        p20.LineStyle='none';
        
        p40=patch(s40);
        p40.FaceColor='g';
        p40.FaceAlpha=0.25;
        p40.LineStyle='none';
        
        p60=patch(s60);
        p60.FaceColor='y';
        p60.FaceAlpha=0.25;
        p60.LineStyle='none';
        
        p80=patch(s80);
        p80.FaceColor='r';
        p80.FaceAlpha=0.25;
        p80.LineStyle='none';
        
        if(exist('reverse','var') && reverse==1)
            set(gca,'Zdir','reverse');
        end
        set(gca,'LineWidth',2.5);
        set(gca,'Box','on');
        set(gca,'BoxStyle','back');
        set(gca,'XGrid','on');
        set(gca,'YGrid','on');
        set(gca,'ZGrid','on');
        set(gca,'FontSize',20);
        set(gca,'FontWeight','bold');
        xlabel('x [\mum]');
        ylabel('y [\mum]');
        zlabel('z [\mum]');
        title('Isosurfaces of collection diagram [% of maximum]');
        axis equal;
        view(75,15);
        legend('10%','20%','40%','60%','80%');
        mTextBox=annotation('textbox');
        mTextBox.FontSize=16;
        mTextBox.Position=[0 0.2 0.1 0.1];
        mTextBox.LineStyle='none';
        stringamax=strcat('Max # of photons=',num2str(maximum));
        stringa10=strcat('V_{10}=',num2str(v10),'{\mu}m^{3}');
        stringa20=strcat('V_{20}=',num2str(v20),'{\mu}m^{3}');
        stringa40=strcat('V_{40}=',num2str(v40),'{\mu}m^{3}');
        stringa60=strcat('V_{60}=',num2str(v60),'{\mu}m^{3}');
        stringa80=strcat('V_{80}=',num2str(v80),'{\mu}m^{3}');
set(mTextBox,'String',{stringamax,stringa10,stringa20,stringa40,stringa60,string
a80});
        
        if(exist('ctrl','var') && ctrl==1)
            figure(2);
            hold on;
            pp10=patch(isosurface(Y,X,Z,(double(vox10)),0.9));
            pp20=patch(isosurface(Y,X,Z,(double(vox20)),0.9));
```

```matlab
            pp40=patch(isosurface(Y,X,Z,(double(vox40)),0.9));
            pp60=patch(isosurface(Y,X,Z,(double(vox60)),0.9));
            pp80=patch(isosurface(Y,X,Z,(double(vox80)),0.9));
            pp10.FaceColor='k';
            pp10.FaceAlpha=0.25;
            pp10.LineStyle='none';

            pp20.FaceColor='b';
            pp20.FaceAlpha=0.25;
            pp20.LineStyle='none';

            pp40.FaceColor='g';
            pp40.FaceAlpha=0.25;
            pp40.LineStyle='none';

            pp60.FaceColor='y';
            pp60.FaceAlpha=0.25;
            pp60.LineStyle='none';

            pp80.FaceColor='r';
            pp80.FaceAlpha=0.25;
            pp80.LineStyle='none';

            if(exist('reverse','var') && reverse==1)
                set(gca,'Zdir','reverse');
            end
            axis equal;
            view(75,15);
        end
    end
end

function output=normalize(input)
%normalize the input matrix between 0 and 1, returning it as output.
%INPUT PARAMETERS
%input: matrix to be normalized.
%OUTPUT VARIABLES
%output: normalized matrix.
    output=(input-min(reshape(input,[],1)))/(max(reshape(input,[],1))-
min(reshape(input,[],1)));
end

function eta=RT_to_eta(matrix,z_length,z_step,r_step,visualize)
%RT_to_eta(matrix,z_length,z_step,y_step) create the matrix of collection
%efficiency from the output of Zemax.
%INPUT PARAMETERS
%matrix: Nx3 matrix containing y (1st column), z (2nd column) and power
%(3rd column) from Zemax.
%z: number of simulation points along the fiber axis.
%z_step: grid step along the fiber axis (in meters).
%r_step: grid step along the radial axis (in meters).
%OUTPUT VARIABLES
%eta: 2D map of collection efficiency.

    [n_rows,~]=size(matrix);
    r_length=n_rows/z_length;
    eta=reshape(matrix(:,3),r_length,z_length)';
    eta=horzcat(fliplr(eta(:,2:end)),eta);

    if visualize==1
        [zl,rl]=size(eta);
        figure(1);
```

```matlab
            imagesc([-(rl-1)/2:(rl-1)/2]*r_step*1e3,[zl-1:-1:0]*z_step*1e3,eta);
            set(gca,'FontSize',20,'FontWeight','bold','LineWidth',2.5);
            colorbar('LineWidth',2.5);
            xlabel('r [mm]','FontWeight','bold');
            ylabel('z [mm]','FontWeight','bold');
            title('Collection efficiency');
            axis equal;
            axis([-0.4 0.4 0 1.2]);
        end
end

function [volume]=volume_reconstruction(halfplane)
    %% Volume reconstruction from the simmetry plane
    [Z,R]=size(halfplane);
    volume=zeros(2*R-1,2*R-1,Z);
    [i1,i2,i3]=ndgrid(1:2*R-1,1:2*R-1,1:Z);
    
    x=i1-R;
    y=i2-R;
    z=i3-1;
    
    [~,r_out,z_out]=cart2pol(x,y,z);
    
    [j1,j2]=meshgrid(1:R,1:Z);
    r_in=j1-1;
    z_in=j2-1;
    
    volume(:)=griddata(r_in,z_in,halfplane,r_out(:),z_out(:),'linear');
    volume(isnan(volume))=0;
    
end

function [top,bench,x,y]=read_grab(filename,n_of_frame,image_size_x,image_size_y,x_step,y_step,gain_top,gain_bench)
%read_grab imports a two-channel grab acquired with ScanImage into Matlab
%workspace. At the same time, it creates two tiff files with the same data,
%one per each channel.
%INPUT PARAMETERS
%filename: name of the file to be read, with extension.
%n_of_frame: number of acquired frame to be averaged.
%image_size_x: dimension, in pixel, of the image along axis x.
%image_size_y: dimension, in pixel, of the image along axis y.
%x_step: spatial sampling step along x.
%y_step: spatial sampling step along y.
%gain_top: gain of channel 1.
%gain_bench: gain of channel 2.
%OUTPUT VARIABLES
%top: image_size x image_size matrix with data from channel 1 (average on
%n_of_frame frames).
%bench: image_size x image_size matrix with data from channel 2 (average on
%n_of_frame frames).
%x: vector containing spatial axis along x.
%y: vector containing spatial axis along y.
%OUTPUT FILES
%top.tiff: image with grab acquired on channel 1.
%bench.tiff: image with grab acquired on channel 2.

    %Definition of spatial output variables
    x=[0:image_size_x-1]*x_step;
    y=[0:image_size_y-1]*y_step;
    
    %Reads the file containing the grab
```

```matlab
    grab=ScanImageTiffReader(filename).data();

    %Splits the two channels and determines frame average and converts in
    %number of photons
    top=double(median(grab(:,:,1:2:2*n_of_frame-1),3)/gain_top); %odd frames
    bench=double(median(grab(:,:,2:2:2*n_of_frame),3)/gain_bench); %even frames

    %Converts data to uint16 for file writing.
    top16=uint16(top);
    bench16=uint16(bench);

    %Writes the output files.
    imwrite(top16,'top.tiff');
    imwrite(bench16,'bench.tiff');

end

function [top,bench,z,x,y]=read_stack(filename,n_of_slice,n_of_frame,image_size_x,image_size_y,z_step,x_step,y_step,gain_top,gain_bench)
%read_stack imports a two-channel stack acquired with ScanImage into Matlab
%workspace. At the same time, it creates two tiff files with the same data,
%one per each channel.
%INPUT PARAMETERS
%filename: name of the file to be read, with extension.
%n_of_slice: number of slices acquired in the stack.
%n_of_frame: number of acquired frame to be averaged.
%image_size_x: dimension, in pixel, of the image along axis x.
%image_size_y: dimension, in pixel, of the image along axis y.
%z_step: spatial sampling step along z.
%x_step: spatial sampling step along x.
%y_step: spatial sampling step along y.
%gain_top: gain of channel 1.
%gain_bench: gain of channel 2.
%OUTPUT VARIABLES
%top: image_size x image_size matrix with data from channel 1 (average on
%n_of_frame frames).
%bench: image_size x image_size matrix with data from channel 2 (average on
%n_of_frame frames).
%z: vector containing spatial axis along z.
%x: vector containing spatial axis along x.
%y: vector containing spatial axis along y.
%OUTPUT FILES
%top.tiff: image with grab acquired on channel 1.
%bench.tiff: image with grab acquired on channel 2.

    %Definition of spatial output variables
    x=[0:image_size_x-1]*x_step;
    y=[0:image_size_y-1]*y_step;
    z=[0:n_of_slice-1]*z_step;

    %Reads the file containing the grab
    stack=double(ScanImageTiffReader(filename).data());

    %Splits the two channels and converts in number of photons
    top_temp=stack(:,:,1:2:2*n_of_slice*n_of_frame-1)/gain_top; %odd frames
    bench_temp=stack(:,:,2:2:2*n_of_slice*n_of_frame)/gain_bench; %even frames

    %Reshapes the matrices for frame average in each slice

top_reshape=reshape(top_temp,image_size_x,image_size_y,n_of_frame,n_of_slice);
```

```matlab
    bench_reshape=reshape(bench_temp,image_size_x,image_size_y,n_of_frame,n_of_slice
);

    %Calculates frame average in each slice
    top=squeeze(mean(top_reshape,3));
    bench=squeeze(mean(bench_reshape,3));

    %Converts data to uint16 for file writing.
    top16=uint16(top);
    bench16=uint16(bench);

    %Writes the output files.
    delete('topstack.tiff','benchstack.tiff');
    for i=1:n_of_slice
        imwrite(top16(:,:,i),'topstack.tiff','writemode','append');
        imwrite(bench16(:,:,i),'benchstack.tiff','writemode','append');
    end

end

function [gain_top,gain_bench,photon_top]=read_gain(filename)
%read_gain determines the gain on two acquisition channels based on Poisson
%statistics. The two channels are acquired in the same file through
%ScanImage.
%INPUT PARAMETERS
%filename: name of the file to be read, with extension.
%OUTPUT VARIABLES
%gain_top: gain of channel 1.
%gain_bench: gain of channel 2.

    top=reshape(double(imread(filename,1)),1,[]);
    bench=reshape(double(imread(filename,2)),1,[]);

    gain_top=var(top)/mean(top);
    gain_bench=var(bench)/mean(bench);

    photon_top=mean(top)/gain_top;

end
```